\documentclass{article}%
\usepackage{amsfonts}
\usepackage{amsmath}
\usepackage{amssymb}
\usepackage{graphicx}
\usepackage{appendix}%
\providecommand{\U}[1]{\protect\rule{.1in}{.1in}}

\begin{document}

\title{Entropic Dynamics approach to Quantum Electrodynamics}
\author{Ariel Caticha\\{\small Department of Physics, University at Albany--SUNY, Albany, NY 12222,
USA}}
\date{}
\maketitle

\begin{abstract}
Entropic dynamics (ED) is a framework that allows one to derive quantum theory
as a Hamilton-Killing flow on the cotangent bundle of a statistical manifold.
These flows are such that they preserve the symplectic and the\ (information)
metric geometries; they explain the linearity of quantum mechanics and the
appearance of complex numbers. In this paper the ED framework is extended to
deal with local gauge symmetries. More specifically, on the basis of maximum
entropy methods and information geometry, for an appropriate choice of ontic
variables and constraints, we derive the quantum electrodynamics of radiation
fields interacting with charged particles.\ As a test that despite its
unorthodox foundation the ED\ approach is empirically successful we derive the
Maxwell equations.

\end{abstract}

\section{Introduction}

With Einstein's successful introduction of the concept of photon in 1905 the
old enigma of the nature of radiation -- is it waves or is it particles? --
was brought back into physics \cite{Klein 1964}. It provided the motivation
and the context for de Broglie's matter waves, for Schr\"{o}dinger's wave
mechanics, and for Bohr's complementarity \cite{Jammer 1966}. It also marks
the beginning of the long history of controversy over conceptual issues
concerning the interpretation of quantum mechanics that endure to this day.

Entropic Dynamics (ED) is a framework that was designed to address those
conceptual issues. It provides a \emph{realist }$\psi$\emph{-epistemic}
interpretation that allows the derivation or reconstruction of the
mathematical formalism of quantum theory on the basis of established
principles of inference that include probability theory, entropic methods, and
information geometry \cite{Caticha 2019} (for a pedagogical review see
\cite{Caticha 2025}). Its signal characteristic is a clear distinction of
those variables that represent the ontology from those variables that
represent the epistemology. The ontic variables, those which represent what is
real, have definite values at all times which, however, are generally unknown
and are, therefore, uncertain. Managing these uncertainties requires the
introduction of epistemic variables -- the probabilities -- and it is the
evolution of these probabilities that constitutes the main subject of the
theory. Thus, ED is quite conservative in that it achieves an ontic/epistemic
clarity that is characteristic of classical theories but, on the other hand,
ED is radically non-classical in that the dynamics is totally relegated to the
epistemic sector. ED is a dynamics of probabilities; there is no ontic dynamics.

An important consequence of an entropic dynamics is that if the dynamical
variables --- the probabilities --- are epistemic then so must be their
canonically conjugate momenta. It may be somewhat surprising, but it is
nevertheless true, that there are no ontic momenta directly associated to
either the particles or the potentials. Those quantities are absent from the
quantum ontology; they emerge only in the classical limit. Thus, ED paints an
ontic picture that is radically different from that proposed by classical
physics. Furthermore, being a special instance of the updating of
probabilities, the dynamics of probabilities is severely constrained by the
requirement that it must be consistent with the rules of entropic and Bayesian
inference. It is this restriction that motivates the name `entropic' dynamics.
Incidentally, it is the interpretation of probabilities as epistemic variables
--- that probabilities represent degrees of rational belief --- that demands a
Bayesian approach.

In this paper the framework of Entropic Dynamics is extended to tackle a local
gauge theory. We shall apply it to derive the paradigmatic example of quantum
electrodynamics (QED). The obvious question immediately arises: since QED is
one of the most successful and most studied theories in physics, why bother?
For two reasons. The first is that questions of interpretation still plague
QED. Here, by reconstructing the formalism of QED, we shall clarify issues
such as `what is ontic?' and `what is epistemic?' The second reason for
tackling QED is to understand how to handle gauge invariance within the
context of an entropic dynamics. The standard approaches to the quantum theory
of gauge fields (see \emph{e.g.}, \cite{Fadeev Slavnov 1980}\cite{Henneaux
Teitelboim 1992}\cite{Weinberg 2015}) rely heavily on first formulating a
classical gauge theory \cite{Dirac 1964} which is subsequently modified by the
imposition of quantization rules. ED follows an altogether different approach.
It skips the prior formulation of a classical theory and the application of
quantization rules and proceeds directly to formulating the quantum theory.
The ED approach to quantum gauge fields is conceptually very different from
those advocated in the past.

We shall confine our attention to the QED of non-relativistic point charges.
This is clearly an idealization but it will allow us to focus our attention on
a number of foundational issues relevant to local gauge theories while
avoiding distracting issues (\emph{e.g.}, particle creation, vacuum
polarization, etc.) of relativistic origin. However, not all relativistic
considerations are irrelevant distractions; some previous work related to the
interplay between gauge and relativity effects can be found in \cite{Ipek et
al 2018}\cite{Ipek Caticha 2020} and other important applications will be left
for future work.

An overview of this paper is as follows. In section 2 with introduce the
microstates that represent the real stuff, the particles and the gauge fields,
recognizing that the representation of the latter is redundant. The symplectic
and information geometry of the epistemic phase space of probabilities and
their conjugate momenta is established in section 3. In section 4 we study
kinematics, namely, we characterize those special congruences of curves ---
the Hamilton-Killing flows --- that are adapted to the geometric structures of
the epistemic phase space.

Entropic Dynamics is the subject of section 5. There we use the method of
maximum entropy to calculate the transition probability for an infinitesimally
short step and introduce the notion of entropic time as a device to keep track
of the accumulation of many such short steps. Then we derive the Hamiltonian
that generates evolution in entropic time. As a first, albeit standard,
application we derive the conservation of charge.

In section 6 we complete the derivation of QED. We discuss the additional
requirements that implement gauge invariance and derive the Gauss constraint.
Then, we derive the Maxwell equations and verify the emergence of the Coulomb
potential. These latter developments yield the familiar results, albeit with a
different interpretation; they serve to reassure us that despite its
unorthodox foundations the ED approach is empirically successful. Section 7
provides a brief summary of the main conclusions including the fact that in
\emph{ED there are no ontic matter waves and there are no ontic photon
particles}.

\section{The ontic microstates}

\label{Gauge invariance}In the ED of nonrelativistic matter the ontic
microstates are assigned properties that are clearly recognized as those of
particles: they have definite positions and follow continuous trajectories. On
the other hand the ontic microstates of the radiation are assigned properties
that are clearly recognized as those of fields: to every point $x$ in space
one assigns a vector potential degree of freedom $A_{a}(x)$.

Our system consists of $N$ particles and a radiation field living in a flat
3-dimensional Euclidean space $\mathcal{X}$ with metric $\delta_{ab}$. The
ontic microstates of the particles are represented by the coordinates of their
positions $z_{n}^{a}$, collectively denoted by\emph{ }$z=\{z_{n}^{a}\}$. (The
index $n$ $=1\ldots N$ labels the particles, and $a=1,2,3$ the three spatial
coordinates.) The $3N$-dimensional ontic configuration space for $N$ particles
is denoted $\mathcal{X}_{N}$, so that $z\in\mathcal{X}_{N}$. The ontic
microstates of the radiation field are represented by vector potential
distributions $A_{a}(x)$ with $x\in\mathcal{X}$. To simplify the notation we
shall write $A_{a}(x)=A_{ax}$. The set of field microstates form the ontic
configuration space $\mathcal{A}$; each field configuration is a point
$A\in\mathcal{A}$. The full ontic configuration space is the Cartesian product
$\mathcal{X}_{N}\mathcal{A}$.

We have already made two important assumptions about the ontology: the
particles' positions and the radiation field have definite values at all times
and they follow continuous trajectories. These two assumptions are in
contradiction with the standard Copenhagen interpretation of quantum mechanics.

A third and central assumption is close in spirit to the classical theory of
gauge fields: the representation of the radiation field by the field $A_{ax}$
is redundant in that a gauge shift to the new field
\begin{equation}
A_{ax}^{\xi}=A_{ax}+\partial_{a}\xi_{x}~,~ \label{GT a}%
\end{equation}
where $\xi_{x}$ is a scalar function, represents the same ontic microstate.
(Notation: $\partial_{a}=\partial/\partial x^{a}$.) Thus, a given ontic
microstate of the radiation field can be represented by any member of a whole
family of fields $A^{\xi}$ generated by different functions $\xi_{x}$.

The continuity of the paths leads to an important simplification:\ the
dynamics can be studied as a sequence of infinitesimally short steps as the
system makes a transition from an initial state $(z,A)$ to a neighboring state
$(z^{\prime},A^{\prime})=$ $(z+\Delta z,A+\Delta A)$. We seek to develop an
entropic dynamics that takes into account that the variables $(z^{\prime
},A^{\prime})$ and $(z^{\prime},A^{\prime\xi})$ represent the same ontic microstate.

\section{The epistemic phase space}

The goal of ED is to predict the positions of particles and the values of
fields on the basis of information that turns out to be incomplete\ and forces
us to deal with the probability distribution $\rho_{t}[z,A]$. The notation
$\rho_{t}[z,A]$ is meant to show that $\rho$ is a function of $t$ and $z$ and
is a functional of $A$. The epistemic configuration space, or
\textquotedblleft e-configuration\textquotedblright\ space, is the statistical
manifold $\mathcal{S}$ of normalized gauge-invariant distributions,
$\rho\lbrack z,A^{\xi}]=\rho\lbrack z,A]$.

The restrictions to normalized probabilities and to gauge invariant
distributions are serious technical inconveniences that will be provisionally
handled by embedding the space $\mathcal{S}$ in the larger space
\begin{equation}
\mathcal{S}^{+}=\left\{  \rho|\text{~}\rho\lbrack z,A]\geq0\right\}
\label{econfig}%
\end{equation}
of unnormalized probabilities with no gauge symmetry. The appropriate
restrictions will be later reintroduced as normalization and gauge constraints
on the \textquotedblleft physical\textquotedblright\ states.

The dynamics of probabilities on $\mathcal{S}^{+}$ is, however, best defined
over the corresponding cotangent bundle $T^{\ast}\mathcal{S}^{+}$, which
constitutes the epistemic phase space (or \textquotedblleft
e-phase\textquotedblright\ space). We shall be interested in trajectories in
e-phase space and this leads us to consider those special curves that are
naturally adapted to the geometry of $T^{\ast}\mathcal{S}^{+}$. In this
section we describe the geometry of $T^{\ast}\mathcal{S}^{+}$ in terms of its
symplectic, metric, and complex structures. This is carried out as a
straightforward extension to the infinite dimensional case of quantum fields
of the corresponding structures that apply to discrete systems \cite{Caticha
2021} and to finite dimensional systems \cite{Caticha 2025}. For a
pedagogical\ background on symplectic and metric geometry see \cite{Schutz
1980}.

First we shall establish some notation. For simplicity the coordinates of a
point on $T^{\ast}\mathcal{S}^{+}$ shall be written $(\rho_{zA},\phi_{zA})$
where $\rho_{zA}=\rho\lbrack z,A]$ are coordinates on the base manifold
$\mathcal{S}^{+}$ and $\phi_{zA}=\phi\lbrack z,A]$ are coordinates on the
cotangent space $T_{zA}^{\ast}\mathcal{S}^{+}$. The vector $\bar{V}$ tangent
to a generic curve parametrized by $\lambda$ is written as
\begin{equation}
\bar{V}=\frac{d}{d\lambda}=\int dzDA\left(  \frac{d\rho_{zA}}{d\lambda}%
\frac{\delta}{\delta\rho_{zA}}+\frac{d\phi_{zA}}{d\lambda}\frac{\delta}%
{\delta\phi_{zA}}\right)  =V^{\alpha zA}\frac{\delta}{\delta X^{\alpha zA}}%
\end{equation}
where $dz$ stands for $d^{3N}z$, $DA$ is an appropriate functional measure of
integration, and where we introduced the discrete index $\alpha=1,2$ to stand
for $\rho$ and $\phi$ respectively,
\begin{equation}
X^{1zA}=\rho_{zA}\quad\text{and}\quad X^{2zA}=\phi_{zA}~.\quad
\end{equation}
It is understood that sums and integrations over repeated indices are carried
out. The derivative of a functional $F[\rho,\phi]$ along $\bar{V}$ is%
\begin{equation}
\frac{dF}{d\lambda}=\int dzDA\left(  \frac{d\rho_{zA}}{d\lambda}\frac{\delta
F}{\delta\rho_{zA}}+\frac{d\phi_{zA}}{d\lambda}\frac{\delta F}{\delta\phi
_{zA}}\right)  =\frac{\delta F}{\delta X^{\alpha x}}\frac{dX^{\alpha x}%
}{d\lambda}=\tilde{\nabla}F[\bar{V}]~,
\end{equation}
where $\tilde{\nabla}$ is the gradient in $T^{\ast}\mathcal{S}^{+}$, that is,
\begin{equation}
\tilde{\nabla}F[\cdot]=\int dzDA\left(  \frac{\delta F}{\delta\rho_{zA}}%
\tilde{\nabla}\rho_{zA}[\cdot]+\frac{\delta F}{\delta\phi_{zA}}\tilde{\nabla
}\phi_{zA}[\cdot]\right)  =\frac{\delta F}{\delta X^{\alpha x}}\tilde{\nabla
}X^{\alpha x}[\cdot]~,
\end{equation}
and%
\begin{equation}
\tilde{\nabla}X^{1zA}=\tilde{\nabla}\rho_{zA}\quad\text{and}\quad\tilde
{\nabla}X^{2zA}=\tilde{\nabla}\phi_{zA}%
\end{equation}
are the basis covectors. The tilde `$\symbol{126}$' is meant to distinguish
the gradient $\tilde{\nabla}$ on $T^{\ast}\mathcal{S}^{+}$ from the gradient
$\nabla$ on $\mathcal{S}^{+}$.

\subsection{Symplectic structure}

The coordinates $\rho_{zA}$ and $\phi_{zA}$ have a clear physical significance
that is explicit for $\rho$, which are the \emph{probabilities} of the ontic
variables and will later, in section (\ref{Hamiltonian}), also become explicit
for $\phi$. These privileged coordinates suggest a natural choice for a
symplectic $2$-form, namely,%

\begin{equation}
\Omega\lbrack\cdot,\cdot]=\int dzDA\,\left(  \tilde{\nabla}\rho_{zA}%
[\cdot]\otimes\tilde{\nabla}\phi_{zA}[\cdot]-\tilde{\nabla}\phi_{zA}%
[\cdot]\otimes\tilde{\nabla}\rho_{zA}[\cdot]\right)  ~, \label{x sympl form a}%
\end{equation}
where $\otimes$ is the tensor product. To find the tensor components of
$\Omega$ we look at the action of $\Omega\lbrack\cdot,\cdot]$ on two generic
gauge invariant vector fields $\bar{V}=d/d\lambda$ and $\bar{U}=d/d\mu$. We
find
\begin{equation}
\Omega\lbrack\bar{V},\bar{U}]=\int dzDA\,\left[  V^{1zA}U^{2zA}-V^{2zA}%
U^{1zA}\right]  =\Omega_{\alpha zA,\beta z^{\prime}A^{\prime}}V^{\alpha
zA}U^{\beta z^{\prime}A^{\prime}}~,~ \label{x sympl form b}%
\end{equation}
and the components of $\Omega$ displayed as a matrix take the familiar form
\begin{equation}
\lbrack\Omega_{\alpha zA,\beta z^{\prime}A^{\prime}}]=%
\begin{bmatrix}
0 & 1\\
-1 & 0
\end{bmatrix}
\delta_{\alpha zA,\beta z^{\prime}A^{\prime}}~, \label{x sympl form c}%
\end{equation}
where
\begin{equation}
\delta_{\alpha zA,\beta z^{\prime}A^{\prime}}=\delta_{\alpha\beta}%
\delta_{zz^{\prime}}\delta_{AA^{\prime}}~,
\end{equation}
where $\delta_{\alpha\beta}$ is a Kronecker $\delta$ symbol, $\delta
_{zz^{\prime}}=\delta(z,z^{\prime})$ is the Dirac $\delta$ function, and
$\delta_{AA^{\prime}}=\delta\lbrack A,A^{\prime}]$ is a Dirac $\delta$ functional.

\subsection{Information geometry}

The simplex of normalized probabilities is a statistical manifold. Its metric
geometry, which turns out to be spherically symmetric, is uniquely
determined\ by information geometry up to a multiplicative overall constant
that fixes the units of length (see \cite{Caticha 2025}). The information
geometry of the embedding e-configuration space $\mathcal{S}^{+}$ of
unnormalized probabilities, eq.(\ref{econfig}), is also spherically symmetric
but it is not uniquely determined. It turns out that without loss of
generality (see \cite{Caticha 2025}) the geometries of $\mathcal{S}^{+}$ and
of the e-phase space $T^{\ast}\mathcal{S}^{+}$ can be assigned the simplest of
the spherically symmetric geometries consistent with information geometry,
namely, we can choose them to be flat. With this choice the length element on
$T^{\ast}\mathcal{S}^{+}$ is
\begin{equation}
\delta\tilde{\ell}^{2}=\int dzDA\left[  \frac{\hbar}{2\rho_{zA}}\delta
\rho_{zA}^{2}+\frac{2\rho_{zA}}{\hbar}\delta\phi_{zA}^{2}\right]  =G_{\alpha
zA,\beta z^{\prime}A^{\prime}}\delta X^{\alpha zA}\delta X^{\beta z^{\prime
}A^{\prime}}~. \label{TS+ metric c}%
\end{equation}
Displaying the components of the metric tensor $G$ in matrix form we find
\begin{equation}
\lbrack G_{zA,z^{\prime}A^{\prime}}]=%
\begin{bmatrix}
\frac{\hbar}{2\rho_{zA}} & 0\\
0 & \frac{2\rho_{zA}}{\hbar}%
\end{bmatrix}
\delta_{zA,z^{\prime}A^{\prime}}\,, \label{TS+ metric d}%
\end{equation}
where $\hbar$ is a constant (eventually identified with Planck's constant)
that can be given a geomeric interpretation: it controls the relative
contributions of $\delta\rho_{x}$ and $\delta\phi_{x}$ to the length of the
infinitesimal vector $(\delta\rho,\delta\phi)$.

\subsection{Complex structure}

The inverse of the metric tensor $G^{-1}$ allows us to raise an index of the
symplectic $\Omega$ form and define a new tensor $J$ with components%
\begin{equation}
J^{\alpha zA}{}_{\beta z^{\prime}A^{\prime}}=-G^{\alpha zA,\gamma
z^{\prime\prime}A^{\prime\prime}}\Omega_{\gamma z^{\prime\prime}%
A^{\prime\prime},\beta z^{\prime}A^{\prime}}~. \label{J tensor a}%
\end{equation}
The negative sign is purely conventional. Written as a matrix $J$ takes the
form
\begin{equation}
\lbrack J^{zA}{}_{z^{\prime}A^{\prime}}]=%
\begin{bmatrix}
0 & -\frac{2\rho_{zA}}{\hbar}\\
\frac{\hbar}{2\rho_{zA}} & 0
\end{bmatrix}
\delta_{zA,z^{\prime}A^{\prime}}~. \label{J tensor b}%
\end{equation}
It is easy to check that%
\begin{equation}
J^{\alpha zA}{}_{\gamma z^{\prime\prime}A^{\prime\prime}}J^{\gamma
z^{\prime\prime}A^{\prime\prime}}{}_{\beta z^{\prime}A^{\prime}}%
=-\delta_{\alpha zA,\beta z^{\prime}A^{\prime}}\mathbf{\quad}\text{or}%
\mathbf{\quad}J^{2}=-\mathbf{1}~, \label{J tensor c}%
\end{equation}
which means that $T^{\ast}\mathcal{S}^{+}$ has a complex structure and
suggests that it might be useful to perform a canonical transformation from
the $(\rho,\phi)$ coordinates to complex coordinates $(\psi,i\hbar\psi^{\ast
})$ where%
\begin{equation}
\psi_{zA}=\rho_{zA}^{1/2}e^{i\phi_{zA}/\hbar}~ \label{psi coords}%
\end{equation}
is the wave functional (or, for simplicity, just the wave \textquotedblleft
function\textquotedblright). This is the geometrical reason for complex
numbers in quantum mechanics: the joint presence of a symplectic and a metric
structure implies a complex structure. Such manifolds are usually called
K\"{a}hler manifolds; the e-phase space $T^{\ast}\mathcal{S}^{+}$ is a very
special K\"{a}hler manifold in that its metric structure is derived from
information geometry. For details of the subtleties of the transformation from
$(\rho,\phi)$ to $(\psi,i\hbar\psi^{\ast})$ see \cite{Caticha 2021}%
\cite{Caticha 2025}.

In wave function coordinates, eqs.(\ref{TS+ metric c}) and (\ref{TS+ metric d}%
), become
\begin{equation}
\delta\tilde{\ell}^{2}=2\hbar\int dzDA\,\delta\psi_{zA}^{\ast}\delta\psi
_{zA}=2\hbar\langle\delta\psi|\delta\psi\rangle~,\quad\lbrack G_{zA,z^{\prime
}A^{\prime}}]=-i%
\begin{bmatrix}
0 & 1\\
1 & 0
\end{bmatrix}
\delta_{zA,z^{\prime}A^{\prime}}~. \label{TS+ metric e}%
\end{equation}
The transformation (\ref{psi coords}) is canonical in the sense that the
components of the symplectic tensor $\Omega\,$, eq.(\ref{x sympl form c}),
remain unchanged. Notice also that, unlike the components of the metric tensor
$G$ in $(\rho,\phi)$ coordinates, eq.(\ref{TS+ metric d}), in wave function
coordinates the components of $G$, eq.(\ref{TS+ metric e}), are constants,
which makes explicit both the flatness of the e-phase space $T^{\ast
}\mathcal{S}^{+}$ and the convenience of wave functions.

\section{Kinematics}

\subsection{Hamilton flows}

Given the symplectic form $\Omega$ there are families of vector fields
$\bar{H}$ that are special in the sense that the Lie derivative of $\Omega$
along $\bar{H}$ vanishes,
\begin{equation}
\pounds _{\bar{H}}\Omega=0~, \label{Lie Omega}%
\end{equation}
that is, the symplectic form $\Omega$ is preserved along the integral curves
generated by $\bar{H}$. These curves, which are naturally adapted to the
symplectic structure of $T^{\ast}\mathcal{S}^{+}$, are such that associated to
the vector field $\bar{H}$ there is a scalar function $\tilde{H}$ such that
the components of $\bar{H}$ are given by Hamilton's equations \cite{Schutz
1980}. In wave function coordinates we have,
\begin{equation}
H^{1zA}=\frac{d\psi_{zA}}{d\tau}=\frac{1}{i\hbar}\frac{\delta\tilde{H}}%
{\delta\psi_{zA}^{\ast}}\quad\text{and}\quad H^{2zA}=i\hbar\frac{d\psi
_{zA}^{\ast}}{d\tau}=-\frac{\delta\tilde{H}}{\partial\psi_{zA}}~,
\label{Hamiltonian flow a}%
\end{equation}
where $\tau$ is the parameter along the curves. This justifies calling
$\bar{H}$ the \emph{Hamiltonian vector field} associated to the
\emph{Hamiltonian function} $\tilde{H}$, and the congruence of these curves is
called a Hamiltonian flow.\ Eventually we shall choose $\tilde{H}$ to be the
actual Hamiltonian that generates a flow in time. At this point, however,
$\tilde{H}$ is some generic Hamiltonian function and there is no implication
that the parameter $\tau$ represents time; for example, $\tilde{H}$ could be
the generator of rotations and $\tau$ the corresponding rotation angle.

The action of $\Omega\lbrack\cdot,\cdot]$ on two Hamiltonian vector fields
$\bar{V}=d/d\lambda$ and $\bar{U}=d/d\mu$ generated respectively by $\tilde
{V}$ and $\tilde{U}$ is
\begin{equation}
\Omega(\bar{V},\bar{U})=\int dzDA\,i\hbar\left(  \frac{d\psi_{zA}}{d\lambda
}\frac{d\psi_{zA}^{\ast}}{d\mu}-\frac{d\psi_{zA}^{\ast}}{d\lambda}\frac
{d\psi_{zA}}{d\mu}\right)  ~,
\end{equation}
which, using (\ref{Hamiltonian flow a}), gives%
\begin{equation}
\Omega(\bar{V},\bar{U})=\int dzDA\frac{1}{i\hbar}\left(  \frac{\delta\tilde
{V}}{\delta\psi_{zA}}\frac{\delta\tilde{U}}{\delta\psi_{zA}^{\ast}}%
-\frac{\delta\tilde{V}}{\delta\psi_{zA}^{\ast}}\frac{\delta\tilde{U}}%
{\delta\psi_{zA}}\right)  \overset{\text{def}}{=}\{\tilde{V},\tilde{U}\}~,
\label{PB a}%
\end{equation}
which is recognized as the Poisson bracket.

\subsection{Normalization}

\label{Normalization}We shall eventually demand that probabilities be
normalized. This is expressed as a constraint,
\begin{equation}
\tilde{N}=0\quad\text{where}\quad\tilde{N}=1-\int dzDA\,\psi_{zA}^{\ast}%
\psi_{zA}~, \label{x N constraint}%
\end{equation}
on the physically relevant states. Since the constraint $\tilde{N}%
=\operatorname*{const}$ must be preserved by time evolution, we shall require
Hamiltonian functions such that
\begin{equation}
\frac{d\tilde{N}}{d\tau}=\{\tilde{N},\tilde{H}\}=0~. \label{x N conservation}%
\end{equation}
Conversely, if $\tilde{N}$ is conserved along the flow generated by $\tilde
{H}$, then $\tilde{H}$ is conserved along the Hamilton flow generated by
$\tilde{N}$ and parametrized by $\nu$,
\begin{equation}
\frac{d\tilde{H}}{d\nu}=\{\tilde{H},\tilde{N}\}=0~, \label{global gauge sym}%
\end{equation}
which means that $\tilde{N}$ generates a symmetry. The flow generated by
$\tilde{N}$ is given by Hamilton's equations,
\begin{equation}
\frac{d\psi_{zA}}{d\nu}=\frac{1}{i\hbar}\frac{\delta\tilde{N}}{\delta\psi
_{zA}^{\ast}}=\frac{i}{\hbar}\psi_{zA}~,
\end{equation}
which is easily integrated to give
\begin{equation}
\psi_{zA}(\nu)=\psi_{zA}(0)e^{i\nu/\hbar}~. \label{rays}%
\end{equation}
The interpretation is as follows. The symmetry generated by $\tilde{N}$ is the
consequence of embedding the e-phase space $T^{\ast}\mathcal{S}$ of normalized
probabilities into the larger embedding space $T^{\ast}\mathcal{S}^{+}$ of
unnormalized probabilities. One consequence of introducing one superfluous
coordinate $\rho$ is to also introduce a superfluous momentum $\phi$. The
extra coordinate $\rho$ is eliminated by imposing the constraint $\tilde{N}%
=0$. The extra momentum $\phi$ is eliminated by declaring it unphysical. This
is achieved by declaring that states along the flow are equivalent, that is,
states $\psi_{zA}(\nu)$ with different $\nu$ lie on the same \textquotedblleft
ray\textquotedblright.

\subsection{Hamilton-Killing flows}

We can further restrict the choice of the Hamiltonian function $\tilde{H}$ by
requiring that, in addition to preserving the symplectic structure $\Omega$ it
must also preserve the metric structure $G$. Thus we seek a Hamiltonian
$\tilde{H}$ that satisfies
\begin{equation}
\pounds _{\bar{H}}\Omega=0\quad\text{and}\quad\pounds _{\bar{H}}G=0~,
\label{HK flows}%
\end{equation}
and generates flows that are simultaneously Hamilton flows and Killing flows.
As shown in \cite{Caticha 2021} (see also \cite{Caticha 2025}) the
Hamilton-Killing flows (or HK flows) are generated by Hamiltonians of the
form
\begin{equation}
\tilde{H}[\psi,\psi^{\ast}]=\int dzDAdz^{\prime}DA^{\prime}\,\psi_{zA}^{\ast
}\hat{H}_{zA,z^{\prime}A^{\prime}}\psi_{z^{\prime}A^{\prime}}+\int dzDA\left(
\,\psi_{zA}^{\ast}\hat{L}_{zA}+\hat{M}_{zA}\psi_{zA}\right)  ~,
\end{equation}
where the kernels $\hat{H}$, $\hat{L}$, and $\hat{M}$ are independent of
$\psi^{\ast}$ and $\psi$. Requiring that the flow also preserve normalization
implies that $\tilde{H}$ must be invariant under the phase shift (\ref{rays}).
Therefore, $\hat{L}_{zA}=0$ and $\hat{M}_{zA}=0$, and $\tilde{H}$ must be
bilinear in $\psi_{zA}$ and $\psi_{zA}^{\ast}$, that is,
\begin{equation}
\tilde{H}[\psi,\psi^{\ast}]=\int dzDAdz^{\prime}DA^{\prime}\,\psi_{zA}^{\ast
}\hat{H}_{zA,z^{\prime}A^{\prime}}\psi_{z^{\prime}A^{\prime}}~.
\label{bilinear hamiltonian}%
\end{equation}
The kernel $\hat{H}_{zA,z^{\prime}A^{\prime}}$ can be conveniently interpreted
as the matrix element,
\begin{equation}
\hat{H}_{zA,z^{\prime}A^{\prime}}=\langle zA|\hat{H}|z^{\prime}A^{\prime
}\rangle~,
\end{equation}
of a self-adjoint operator $\hat{H}$. (For a pedagogical presentation of the
Schr\"{o}dinger representation in quantum field theory see \cite{Jackiw
1990}.) Such bilinear Hamiltonians lead to equations of motion,
\begin{equation}
i\hbar\frac{d\psi_{zA}}{d\tau}=\frac{\delta\tilde{H}}{\delta\psi_{zA}^{\ast}%
}=\int dz^{\prime}DA^{\prime}\hat{H}_{zA,z^{\prime}A^{\prime}}\psi_{z^{\prime
}A^{\prime}}~, \label{Sch a}%
\end{equation}
that are recognized as Schr\"{o}dinger equations. (At this point in the
argument there is no implication yet that the parameter $\tau$ represents
time.) Thus, the requirement of HK flows that also preserve normalization
provides an explanation for the linearity of quantum mechanics and its
attendant superposition principle.

The main result of this section is that the special curves that are adapted to
the symplectic and the information metric structures of the e-phase $T^{\ast
}\mathcal{S}^{+}$ are HK-flows generated by bilinear Hamiltonians. This
concludes our discussion of kinematics.

\section{Entropic dynamics}

Having figured that HK-flows yield specially privileged families of curves our
next goal is to figure out which of those curves are good candidates to be
actual trajectories. We want to derive the dynamics of probabilities. First,
we use the method of maximum entropy to calculate the transition probability
of a short step; this is the central idea behind any \emph{entropic} dynamics.
Then, since the rules of inference are atemporal, the assumptions behind the
construction of the concept of time must be explicitly stated. An
\emph{entropic} notion of time is introduced as a book-keeping device to keep
track of the accumulation of a succession of short steps. The final step
yields a Hamiltonian that generates an HK-flow in entropic time and allows us
to write the Schr\"{o}dinger functional equation. Similar ideas have been
previously developed in \cite{Caticha 2021}\cite{Caticha 2025}. Here the ED
framework will be further extended to the case of a field with a local gauge symmetry.

\subsection{The transition probability for a short step}

\label{transition probability}The continuity of trajectories in the ontic
configuration space allows us to analyze the evolution as a sequence of short
steps, which leads us to calculate the transition probability $P[z^{\prime
}A^{\prime}|zA]=P_{z^{\prime}A^{\prime}|zA}$ from $z$ to $z^{\prime}=z+\Delta
z$ and from $A$ to $A^{\prime}=A+\Delta A$. To guarantee that the dynamics is
consistent with the rules for updating probabilities the transition
probability $P_{z^{\prime}\hat{A}^{\prime}|zA}$ is found by maximizing the
entropy,
\begin{equation}
\mathcal{S}[z^{\prime}A^{\prime}|zA]=-\int dz^{\prime}DA^{\prime
}\,P_{z^{\prime}A^{\prime}|zA}\log\frac{P_{z^{\prime}A^{\prime}|zA}%
}{Q_{z^{\prime}A^{\prime}|zA}}~. \label{entropy}%
\end{equation}
relative to the prior $Q$ and subject to the appropriate constraints. The
relevant physical information is introduced in two different ways. The
information that is common to all short steps is codified into the prior
$Q_{z^{\prime}A^{\prime}|zA}$ while the information that is specific to each
individual short step is introduced as a constraint on the transition
probability $P_{z^{\prime}A^{\prime}|zA}$.

The prior $Q$ codifies the information that the particles and the field evolve
by taking infinitesimally short steps but $Q$ is otherwise maximally
uninformative in the sense that it must reflect the translational and
rotational invariance of the space $\mathcal{X}$ and it must express total
ignorance about any correlations. The desired prior is Gaussian,
\begin{equation}
Q_{z^{\prime}A^{\prime}|zA}\propto\exp-\frac{1}{2}\left[
{\displaystyle\sum\nolimits_{n}}
\alpha_{n}\delta_{ab}\Delta z_{n}^{a}\Delta z_{n}^{b}+\alpha\int
d^{3}x\,\delta^{ab}\Delta A_{ax}\Delta A_{bx}\right]  ~. \label{prior}%
\end{equation}
where the constants $\alpha_{n}$ and $\alpha$ are Lagrange multipliers to be
specified later. (The Gaussian prior $Q$ can itself be derived using the
method of maximum entropy.) The multipliers $\alpha_{n}$ and $\alpha$ will
eventually be made arbitrarily large to ensure that $Q$ leads to intrinsic
changes that are infinitesimally small. The index $n$ in $\alpha_{n}$
indicates that the particles need not be identical.

The physical information about directionality and correlations is specific to
each short step. It is introduced via a constraint that involves a functional
$\varphi\lbrack z,A]=\varphi_{zA}$ called the \textquotedblleft drift
potential.\textquotedblright\ The constraint consists of imposing that the
expected change of $\varphi_{zA}$ over a short step is some small amount
$\kappa^{\prime}$,%
\begin{equation}
\langle\Delta\varphi_{zA}\rangle=\int dz^{\prime}DA^{\prime}\,P_{z^{\prime
}A^{\prime}|zA}\,\Delta\varphi_{zA}=\kappa^{\prime}~.
\end{equation}
More explicitly, we can Taylor expand $\Delta\varphi_{zA}$ and write
\begin{equation}
\int dz^{\prime}DA^{\prime}\,P_{z^{\prime}A^{\prime}|zA}\left[
{\displaystyle\sum\nolimits_{n}}
\frac{\partial\varphi_{zA}}{\partial z_{n}^{a}}\Delta z_{n}^{a}+\int
d^{3}x\,\frac{\delta\varphi_{zA}}{\delta A_{ax}}\Delta A_{ax}\right]
=\kappa^{\prime}~. \label{drift constraint}%
\end{equation}
As in other forms of ED, the drift potential $\varphi_{zA}$ is of central
importance. It plays three separate roles: (1) it serves to express a
constraint that codifies physical information; (2) it turns out to be a
momentum that is canonically conjugate to the probability $\rho_{zA}$; and (3)
it will be closely related to the phase $\phi_{zA}$ of the wave functional
$\psi_{zA}$ in eq.(\ref{psi coords}).

Finally, we require one last set of constraints that describes how each
particle interacts with the gauge field $A$. We impose that the expected
displacement $\langle\Delta z_{n}^{a}\rangle$ for each particle satisfies
\begin{equation}
\langle\Delta z_{n}^{a}\rangle A_{az_{n}}=\kappa_{n}^{\prime\prime}%
\quad\text{for\quad}n=1\ldots N~. \label{gauge constraints}%
\end{equation}
The coupling is local: the particle at $z$ couples to the field at $z$. The
numerical values of the quantities $\kappa^{\prime}$ and $\kappa_{n}%
^{\prime\prime}$ could be specified directly but it is more convenient to
specify them indirectly in terms of the corresponding Lagrange multipliers.

We are now ready to assign the transition probability $P_{z^{\prime}A^{\prime
}|zA}$. Maximize the entropy (\ref{entropy}) relative to the prior
(\ref{prior}) and subject to the constraints (\ref{drift constraint}),
(\ref{gauge constraints}), and normalization. The result is a Gaussian
distribution,
\begin{align}
P_{z^{\prime}A^{\prime}|zA}  &  \propto\exp%
{\displaystyle\sum\nolimits_{n}}
\left(  -\frac{\alpha_{n}}{2}\delta_{ab}\Delta z_{n}^{a}\Delta z_{n}%
^{b}+\alpha^{\prime}\left(  \partial_{na}\varphi_{zA}-\beta_{n}A_{az_{n}%
}\right)  \Delta z_{n}^{a}\right) \nonumber\\
&  \times\exp\int d^{3}x\,\left(  -\frac{\alpha}{2}\delta^{ab}\Delta
A_{ax}\Delta A_{bx}+\alpha^{\prime}\frac{\delta\varphi_{zA}}{\delta A_{ax}%
}\Delta A_{ax}\right)  ~, \label{tr prob a}%
\end{align}
where $\alpha_{n}$, $\alpha$, $\alpha^{\prime}$, and $\alpha^{\prime}\beta
_{n}$ are Lagrange multipliers, and $\partial_{na}=\partial/\partial z_{n}%
^{a}$. Later we will write the $\beta_{n}$ parameters as $\beta_{n}=q_{n}/c$
where the $q_{n}$ will be identified with the electric charges in
Heaviside-Lorentz units (\emph{i.e.}, rationalized Gaussian units) and $c$ is
the speed of light. Completing the square $P$ can be rewriten as
\begin{align}
P_{z^{\prime}A^{\prime}|zA}  &  \propto\exp-\frac{1}{2}%
{\displaystyle\sum\nolimits_{n}}
\alpha_{n}\delta_{ab}\left(  \Delta z_{n}^{a}-\overline{\Delta z}_{n}%
^{a}\right)  \left(  \Delta z_{n}^{b}-\overline{\Delta z}_{n}^{b}\right)
\nonumber\\
&  \left.  +\int d^{3}x\,\alpha\delta^{ab}\left(  \Delta A_{ax}-\overline
{\Delta A}_{ax}\right)  \left(  \Delta A_{bx}-\overline{\Delta A}_{bx}\right)
\right]  \label{tr prob b}%
\end{align}
where the expected steps $\overline{\Delta z}_{n}^{a}$ and $\overline{\Delta
A}_{ax}$ are
\begin{equation}
\langle\Delta z_{n}^{a}\rangle=\overline{\Delta z}_{n}^{a}=\frac
{\alpha^{\prime}}{\alpha_{n}}\delta^{ab}\left(  \partial_{nb}\varphi
_{zA}-\beta_{n}A_{bz_{n}}\right)  ~, \label{short step z}%
\end{equation}%
\begin{equation}
\langle\Delta A_{ax}\rangle=\overline{\Delta A}_{ax}=\frac{\alpha^{\prime}%
}{\alpha}\delta_{ab}\frac{\delta\varphi_{zA}}{\delta A_{bx}}~.
\label{short step A}%
\end{equation}
Let a generic step be described as
\begin{equation}
\Delta z_{n}^{a}=\overline{\Delta z}_{n}^{a}+\Delta w_{n}^{a}\quad
\text{and}\quad\Delta A_{ax}=\overline{\Delta A}_{ax}+\Delta W_{ax}~,
\label{short step zA}%
\end{equation}
then the fluctuations $\Delta w_{n}^{a}$ and $\Delta W_{ax}$ satisfy,%
\begin{align}
\langle\Delta w_{n}^{a}\rangle &  =0\,\quad\text{and}\hspace{0.4cm}%
\langle\Delta w_{n}^{a}\Delta w_{n^{\prime}}^{b}\rangle=\frac{1}{\alpha_{n}%
}\delta_{ab}\delta_{nn^{\prime}}~,\label{short step c}\\
\langle\Delta W_{ax}\rangle &  =0\quad\text{and}\hspace{0.4cm}\langle\Delta
W_{ax}\Delta W_{bx^{\prime}}\rangle=\frac{1}{\alpha}\delta_{ab}\delta
_{xx^{\prime}}~, \label{short step d}%
\end{align}
while the correlations vanish,
\begin{equation}
\langle\Delta z_{n}^{a}\Delta A_{bx}\rangle=\langle\Delta z_{n}^{a}%
\rangle\langle\Delta A_{bx}\rangle\quad\text{and}\hspace{0.4cm}\langle\Delta
w_{n}^{a}\Delta W_{bx}\rangle=\langle\Delta w_{n}^{a}\rangle\langle\Delta
W_{bx}\rangle=0~. \label{short step e}%
\end{equation}

\subsection{Entropic time}

An entropic dynamics of probabilities requires an entropic notion of time.
This involves introducing the concept of instants that are suitably ordered,
and adopting a convenient measure of the interval between instants. In ED an
instant is specified by the information --- codified into the functions
$\rho_{zA}$ and $\varphi_{zA}$ --- that is sufficient for generating the next
instant. Indeed, if the distribution $\rho$ and the drift potential $\varphi$
refer to the instant $t$, then the distribution
\begin{equation}
\rho_{z^{\prime}A^{\prime}}^{\prime}=\int dzDA\,P_{z^{\prime}A^{\prime}%
|zA}\rho_{zA} \label{DefTime}%
\end{equation}
generated from $\rho_{zA}$ by $P_{z^{\prime}\hat{A}^{\prime}|zA}$ serves both
to define the dynamics of $\rho$ and to construct what we mean by the
\textquotedblleft next\textquotedblright\ instant $t^{\prime}$. Later we shall
complete the construction by specifying the evolution of $\varphi_{zA}$.

Before proceeding further a couple of remarks are in order. In
eq.(\ref{DefTime}) we see that given the information codified into a present
instant $t$, the future instant $t^{\prime}>t$ turns out to be independent of
any past instant $t^{\prime\prime}<t$. Thus, formally, ED is a Markovian
process. However, it is important to emphasize that there is an important
difference. Markovian processes are assumed to develop in an already existing
time determined by external clocks. In ED there is no such pre-existing time;
information about the system at one instant is used to generate or construct
the next instant. Thus, ED is fully relational with respect to time
\cite{Caticha Saleem 2025}.

The second remark is that the evolution equation involves a transition
probability that was derived by maximizing an entropy and therefore there is a
clear evolution from a past instant towards a future instant. In ED the
instants are therefore ordered and there is a natural arrow of time.

We shall now address the unfinished task of specifying the Lagrange
multipliers that appear in the transition probability, eq.(\ref{tr prob a}).
As discussed in \cite{Caticha 2025} the last ingredient in the construction of
time is specifying the measure of duration--- the interval $\Delta t$ between
successive instants --- in terms of the multipliers $\alpha_{n}$, $\alpha$,
and $\alpha^{\prime}$. Thus, \emph{the transition probability provides us with
a clock}. As a matter of convenience we choose multipliers adapted to the
translational symmetries of Newtonian space and, accordingly, we shall choose
them to be constants independent of $x$ and $t\,$.

Furthermore, we want the transition probability to lead to infinitesimally
short steps and particles with well-defined expected velocities.
Eq.(\ref{short step z}) shows that this is achieved choosing the ratio
$\alpha^{\prime}/\alpha_{n}\ $proportional to $\Delta t$,
\begin{equation}
\frac{\alpha^{\prime}}{\alpha_{n}}=\frac{1}{m_{n}}\Delta t~,
\label{alpha ratio a}%
\end{equation}
where we introduced proportionality constants $m_{n}$ that will eventually be
identified with the particle masses. As in previous work we shall also take
that $\alpha^{\prime}=$ $1/\eta$ to be constant and choose $\eta$ so that if
$\Delta t$ has units of time, then $m_{n}$ has units of mass. Then,
\begin{equation}
\alpha^{\prime}=\frac{1}{\eta}\quad\text{so that\quad}\alpha_{n}=\frac{m_{n}%
}{\eta\Delta t}~. \label{alpha prime}%
\end{equation}
The expected drift for particles, eq.(\ref{short step z}), and their
fluctuations, eq.(\ref{short step c}), become%
\begin{equation}
\overline{\Delta z}_{n}^{a}=\frac{\Delta t}{m_{n}}\delta^{ab}\left(
\partial_{nb}\varphi_{zA}-\beta_{n}A_{bz_{n}}\right)  \quad\text{and}%
\quad\langle\Delta w_{n}^{a}\Delta w_{n^{\prime}}^{b}\rangle=\Delta
t\frac{\eta}{m_{n}}\delta^{ab}\delta_{nn^{\prime}}~. \label{drift fluct a}%
\end{equation}
As far as the radiation goes, its multiplier $\alpha$ is also chosen to lead
to a well-defined expected rate of change,
\begin{equation}
\alpha=\frac{1}{c^{2}\eta\Delta t}~.
\end{equation}
The proportionality constant $\alpha$ is such that the field $A$ is measured
in Heaviside-Lorentz units. Then,
\begin{equation}
\overline{\Delta A}_{ax}=\Delta t\,c^{2}\delta_{ab}\frac{\delta\varphi_{zA}%
}{\delta A_{bx}}\quad\text{and}\quad\langle\Delta W_{ax}\Delta W_{bx^{\prime}%
}\rangle=\Delta t\,c^{2}\eta\delta_{ab}\delta_{xx^{\prime}}~.
\label{drift fluct b}%
\end{equation}

\subsection{The Hamiltonian}

\label{Hamiltonian}Our next goal is to identify a Hamiltonian $\tilde{H}%
[\rho,\phi]$\ that generates evolution in entropic time. The traditional
approach in mechanics makes use of a Lagrangian $L(q,\dot{q})$ to identify the
momentum $p=\partial L/\partial\dot{q}$ that is conjugate to the coordinate
$q$. In ED\ no Lagrangians are available and an alternative recipe is
required.\ We propose that a momentum conjugate to $\rho_{zA}$ can be
identified from eq.(\ref{DefTime}) which is the evolution equation for $\rho$
written in integral form. The process takes a few steps. The first step
consists in rewriting eq.(\ref{DefTime}) in differential form,
\begin{equation}
\frac{\partial\rho_{zA}}{\partial t}=-%
{\displaystyle\sum\nolimits_{n}}
\frac{\partial}{\partial z_{n}^{a}}\left(  v_{n}^{a}\rho_{zA}\right)  -\int
d^{3}x\,\frac{\delta}{\delta A_{ax}}\left(  u_{ax}\rho_{zA}\right)  ~.
\label{rho dot a}%
\end{equation}
This is a continuity equation where the velocity of the probability flow ---
the current velocity --- has components along $z_{n}^{a}$ and along $A_{ax}$
given by
\begin{equation}
v_{n}^{a}[z,A]=\frac{\delta^{ab}}{m_{n}}\left(  \partial_{nb}\phi_{zA}%
-\beta_{n}A_{bz_{n}}\right)  \quad\text{and}\quad u_{ax}[z,A]=c^{2}\delta
_{ab}\frac{\delta\phi_{zA}}{\delta A_{bx}}~, \label{curr velocity}%
\end{equation}
where
\begin{equation}
\phi_{zA}=\varphi_{zA}-\eta\log\rho_{zA}^{1/2}~.~ \label{phase}%
\end{equation}
The derivation leading from (\ref{DefTime}) to (\ref{rho dot a}) is given in
the Appendix. The second step is to show that (\ref{rho dot a}) can be written
in Hamiltonian form,
\begin{equation}
\frac{\partial\rho_{zA}}{\partial t}=\frac{\delta\tilde{H}}{\delta\phi_{zA}}~,
\label{rho dot b}%
\end{equation}
which suggests $\phi_{zA}$ is a convenient canonical momentum. Equation
(\ref{phase}) shows that $\phi_{zA}$ and $\varphi_{zA}$ differ by a function
of the generalized coordinate $\rho_{zA}$.\ This is just a canonical
transformation and the choice of either $\varphi_{zA}$ or $\phi_{zA}$ as the
conjugate momentum is just a matter of convenience. To show that
(\ref{rho dot a}) can be written in the form (\ref{rho dot b})\ we must find
the Hamiltonian $\tilde{H}$. This is straightforward: equating
(\ref{rho dot a}) to (\ref{rho dot b}) yields a linear functional differential
equation that can be easily integrated for $\tilde{H}$. The result is,
\begin{align}
\tilde{H}[\rho,\phi]  &  =\int dzDA\,\rho_{zA}\left[
{\textstyle\sum\limits_{n}}
\frac{\delta^{ab}}{2m_{n}}\left(  \partial_{na}\phi_{zA}-\beta_{n}A_{az_{n}%
}\right)  \left(  \partial_{nb}\phi_{zA}-\beta_{n}A_{bz_{n}}\right)  \right.
\nonumber\\
&  +\int d^{3}x\,\frac{c^{2}}{2}\delta_{ab}\frac{\delta\phi_{zA}}{\delta
A_{ax}}\frac{\delta\phi_{zA}}{\delta A_{bx}}+\tilde{F}[\rho]
\label{Hamiltonian a}%
\end{align}
where the unspecified functional $\tilde{F}[\rho]$ is an integration constant.
We write $\tilde{F}[\rho]$ to emphasize that $\tilde{F}$ is independent of
$\phi$, but in principle we could have $\tilde{F}=\tilde{F}[\rho;z,A,t]$. It
is straightforward to check that (\ref{Hamiltonian a}) is a solution of
(\ref{rho dot b}): just take a variation $\delta\phi_{zA}$ and integrate by parts.

Yet another piece of information that guides our choice of $\tilde{H}$ is that
in order to generate HK flows it must take the bilinear form,
eq.(\ref{bilinear hamiltonian}).\ Transforming to wave function coordinates
eq.(\ref{Hamiltonian a}) becomes
\begin{equation}
\tilde{H}=\int dzDA\,\psi_{zA}^{\ast}\left(
{\textstyle\sum\limits_{n}}
\frac{-\hbar^{2}}{2m_{n}}\delta^{ab}D_{na}D_{nb}-\int d^{3}x\,\frac{(\hbar
c)^{2}}{2}\delta_{ab}\frac{\delta^{2}}{\delta A_{ax}\delta A_{bx}}\right)
\psi_{zA}+\tilde{V}[\rho] \label{Hamiltonian b}%
\end{equation}
where, setting $\beta_{n}=q_{n}/c$, $D_{na}$ is the covariant derivative,
\begin{equation}
D_{na}=\frac{\partial}{\partial z_{n}^{a}}-i\frac{q_{n}}{\hbar c}A_{a}%
(z_{n})~, \label{cov deriv}%
\end{equation}
and the new functional $\tilde{V}[\rho]$ differs from the old $\tilde{F}%
[\rho]$ by a functional of $\rho$,
\begin{equation}
\tilde{V}[\rho]=\tilde{F}[\rho]-\int dzDA\,\left(
{\textstyle\sum\limits_{n}}
\frac{\hbar^{2}}{8m_{n}}\frac{\delta^{ab}}{\rho_{zA}}\frac{\partial\rho_{zA}%
}{\partial z_{n}^{a}}\frac{\partial\rho_{zA}}{\partial z_{n}^{b}}+\int
d^{3}x\,\frac{(\hbar c)^{2}}{8}\frac{\delta_{ab}}{\rho_{zA}}\frac{\delta
\rho_{zA}}{\delta A_{ax}}\frac{\delta\rho_{zA}}{\delta A_{bx}}\right)  ~.
\end{equation}
In order to generate an HK\ flow the functional $\tilde{V}[\rho]$ must itself
be bilinear,
\begin{equation}
\tilde{V}[\psi,\psi^{\ast}]=\int dz^{\prime}DA^{\prime}dz^{\prime\prime
}DA^{\prime\prime g}\,\psi_{z^{\prime}A^{\prime}}^{\ast}\hat{V}_{z^{\prime
}A^{\prime},z^{\prime\prime}A^{\prime\prime}}\psi_{z^{\prime\prime}%
A^{\prime\prime}}~, \label{V a}%
\end{equation}
for some Hermitian kernel $\hat{V}$, while still remaining independent of
$\phi$,
\begin{equation}
\frac{\delta\tilde{V}[\psi,\psi^{\ast}]}{\delta\phi_{zA}}=\frac{\delta
\tilde{V}[\rho]}{\delta\phi_{zA}}=0~. \label{V b}%
\end{equation}
Substituting (\ref{V a}) into (\ref{V b}) leads to
\begin{equation}
\int dz^{\prime}DA^{\prime}dz^{\prime\prime}DA^{\prime\prime}\,\left(
-\delta_{z^{\prime}A^{\prime},zA}+\delta_{z^{\prime\prime}A^{\prime\prime}%
,zA}\right)  \psi_{z^{\prime}A^{\prime}}^{\ast}\hat{V}_{z^{\prime}A^{\prime
},z^{\prime\prime}A^{\prime\prime}}\psi_{z^{\prime\prime}A^{\prime\prime}}=0~,
\end{equation}
which must be satisfied for all choices of $\psi_{z^{\prime}A^{\prime}}^{\ast
}$ and $\psi_{z^{\prime\prime}A^{\prime\prime}}$. Therefore, the kernel
$\hat{V}_{z^{\prime}A^{\prime},z^{\prime\prime}A^{\prime\prime}}$ must be
local in the ontic configuration space $\mathcal{X}_{N}\mathcal{A}$,
\begin{equation}
\hat{V}_{z^{\prime}A^{\prime},z^{\prime\prime}A^{\prime\prime}}=\delta
_{z^{\prime}A^{\prime},z^{\prime\prime}A^{\prime\prime}}\hat{V}_{z^{\prime
}A^{\prime}}~, \label{V c}%
\end{equation}
where $\hat{V}_{zA}$ is a real functional that plays the role of a potential.
This is how potentials are introduced in the ED\ framework. Substituting
(\ref{V c}) into (\ref{Hamiltonian b}) we obtain the Hamiltonian
\begin{equation}
\tilde{H}=\int dzDA\,\psi_{zA}^{\ast}\hat{H}\psi_{zA}\,, \label{Hamiltonian c}%
\end{equation}
where the kernel $\hat{H}$ is the operator
\begin{equation}
\hat{H}\overset{\text{def}}{=}%
{\textstyle\sum\limits_{n}}
\frac{-\hbar^{2}}{2m_{n}}\delta^{ab}D_{na}D_{nb}+\int d^{3}x\,\frac{(\hbar
c)^{2}}{2}\delta_{ab}\frac{\delta^{2}}{\delta A_{ax}\delta A_{bx}}+\hat
{V}_{zA}\,. \label{Hamiltonian d}%
\end{equation}
For brevity we will refer to both $\tilde{H}$ and $\hat{H}$ as
\textquotedblleft the Hamiltonian\textquotedblright. The Schr\"{o}dinger
equation is the Hamilton equation of motion, eq.(\ref{Sch a}),
\begin{equation}
\frac{d\psi_{zA}}{dt}=\frac{1}{i\hbar}\frac{\delta\tilde{H}}{\delta\psi
_{zA}^{\ast}}\quad\text{or}\quad i\hbar\frac{d\psi_{zA}}{dt}=\hat{H}_{zA}%
\psi_{zA}~. \label{Sch b}%
\end{equation}
A familiar example is the potential $\hat{V}_{zA}$ for the pure
electrodynamics of radiation and particles in vacuum,%
\begin{equation}
\hat{V}_{zA}=\int d^{3}x\,\frac{1}{2}\delta_{ab}\hat{B}_{x}^{a}\hat{B}_{x}%
^{b}\quad\text{where}\quad\vec{B}_{x}\overset{\text{def}}{=}\vec{\partial}%
_{x}\times\vec{A}_{x}~, \label{potential}%
\end{equation}
Additional terms may be included to describe direct particle-particle
interactions and the polarization effects or the index of refraction in a
material medium.

\subsection{Charge conservation}

At this point in the argument we do not yet have a locally gauge invariant
dynamics; this will depend on appropriate choices of the potential\ $\hat
{V}_{zA}$ and of the initial wave function $\psi(t_{0})$. However,
irrespective of the choice of\ $\hat{V}_{zA}$ and $\psi(t_{0})$, there is a
symmetry under the global phase transformation
\begin{equation}
\psi_{zA}\rightarrow\psi_{zA}^{\lambda}=\psi_{zA}e^{i\lambda}\quad
\text{and}\quad A_{ax}\rightarrow A_{ax}^{\lambda}=A_{ax}%
\end{equation}
where $\lambda$ is a constant. One expects that there is a conservation law
associated to this symmetry and, indeed, there is. The derivation can be
carried out following any of several standard methods. We shall make use
Hamilton's equation (\ref{rho dot b}) which is the continuity equation
(\ref{rho dot a}).

The operators that represent the densities of electric charge and electric
current are
\begin{equation}
\hat{\rho}_{x}^{\text{e}}[z,A]=%
{\displaystyle\sum\nolimits_{n}}
q_{n}\delta_{z_{n}x}\quad\text{and}\quad\hat{J}_{x}^{\text{e}a}[z,A]=%
{\displaystyle\sum\nolimits_{n}}
q_{n}\delta_{z_{n}x}v_{n}^{a}~. \label{densities}%
\end{equation}
The corresponding Hamiltonian functionals are
\begin{equation}
\tilde{\rho}_{x}^{\text{e}}=\int dzDA\,\psi_{zA}^{\ast}\hat{\rho}%
_{x}^{\text{e}}\psi_{zA}=\int dzDA\,\rho_{zA}%
{\displaystyle\sum\nolimits_{n}}
q_{n}\delta_{z_{n}x}~, \label{dens charge}%
\end{equation}
and
\begin{equation}
\tilde{J}_{x}^{\text{e}a}=\int dzDA\,\psi_{zA}^{\ast}\hat{J}_{x}^{\text{e}%
a}\psi_{zA}=\int dzDA\,\rho_{zA}%
{\displaystyle\sum\nolimits_{n}}
q_{n}\delta_{z_{n}x}v_{n}^{a}~. \label{dens current}%
\end{equation}
Using eq.(\ref{rho dot a}) we find%
\begin{align*}
\frac{\partial\tilde{\rho}_{x}^{\text{e}}}{\partial t}  &  =\int
dzDA\,\hat{\rho}_{x}^{\text{e}}\frac{\partial\rho_{zA}}{\partial t}=\\
&  =-\int dzDA%
{\displaystyle\sum\nolimits_{n^{\prime}}}
q_{n^{\prime}}\delta_{z_{n^{\prime}}x}\left(
{\displaystyle\sum\nolimits_{n}}
\frac{\partial}{\partial z_{n}^{a}}\left(  v_{n}^{a}\rho_{zA}\right)  +\int
d^{3}x\,\frac{\delta}{\delta A_{ax}}\left(  u_{ax}\rho_{zA}\right)  \right)
\end{align*}
The integral over $DA$ of the last term is a vanishing surface term,%
\begin{equation}
\int DA\,\frac{\delta}{\delta A_{ax}}\left(  u_{ax}\rho_{zA}\right)  =0~,
\label{intDA}%
\end{equation}
therefore%
\[
\frac{\partial\tilde{\rho}_{x}^{\text{e}}}{\partial t}=-\int dzDA%
{\displaystyle\sum\nolimits_{n^{\prime}n}}
q_{n^{\prime}}\delta_{z_{n^{\prime}}x}\frac{\partial}{\partial z_{n}^{a}%
}\left(  v_{n}^{a}\rho_{zA}\right)  ~.
\]
Integrating by parts and using
\begin{equation}%
{\displaystyle\sum\nolimits_{n^{\prime}}}
q_{n^{\prime}}\frac{\partial}{\partial z_{n}^{a}}\delta(\vec{z}_{n^{\prime}%
}-\vec{x})=q_{n}\frac{\partial}{\partial z_{n}^{a}}\delta(\vec{z}_{n}-\vec
{x})=-q_{n}\frac{\partial}{\partial x^{a}}\delta(\vec{z}_{n}-\vec{x})~,
\end{equation}
we find
\[
\frac{\partial\tilde{\rho}_{x}^{\text{e}}}{\partial t}=-\frac{\partial
}{\partial x^{a}}\int dzDA\,%
{\displaystyle\sum\nolimits_{n}}
q_{n}\delta_{z_{n}x}v_{n}^{a}\rho_{zA}~,
\]
which, using (\ref{dens current}), yields the desired result,
\begin{equation}
\frac{\partial}{\partial t}\tilde{\rho}_{x}^{\text{e}}=-\frac{\partial
}{\partial x^{a}}\tilde{J}_{x}^{\text{e}a}~. \label{charge conserv}%
\end{equation}

We have derived a perfectly legitimate conservation law but there is a
peculiar feature that merits a comment. First, as is evident from
(\ref{dens charge}) and (\ref{dens current}), $\tilde{\rho}_{x}^{\text{e}}$
and $\tilde{J}_{x}^{\text{e}a}$ are expected values. We claim we have a charge
conservation law because the identity (\ref{charge conserv}) holds for any
arbitrary choice of the probability $\rho\lbrack z,A]$, but the electric
charges $q_{n}$ are not some ontic substance that is physically attached to
point particles but parameters that operate at the epistemic level of
probabilities. They are Lagrange multipliers that regulate the probabilistic
correlations between particles and fields. Indeed, as befits an entropic
dynamics of probabilities, the velocity $v_{n}^{a}[z,A]$ that defines the
electric current is not the velocity of the particles but the velocity of the
probability flow.

So far we have discussed ED on the larger unconstrained e-phase space
$T^{\ast}\mathcal{S}^{+}$. Next we discuss the additional requirements that
implement gauge invariance and yield a proper entropic electrodynamics.

\section{Quantum electrodynamics}

\subsection{Gauge invariance and the Gauss constraint}

\label{Gauss Constraint}We require that probabilities, phases, and wave
functions reflect the consequences of invariance under the local gauge
transformation
\begin{equation}
A_{ax}\rightarrow A_{ax}^{\xi}=A_{ax}+\partial\xi_{x}/\partial x^{a}~.
\label{GT b}%
\end{equation}
The main constraint follows from the observation that $A$ and $A^{\xi}$
represent the same ontic microstate and therefore the corresponding
probabilities must be equal,
\begin{equation}
\rho\lbrack z,A^{\xi}]\approx\rho\lbrack z,A]~. \label{GI a}%
\end{equation}
(We adopt Dirac's weak equality $\approx$ notation that holds once the
constraints are implemented.) This can be expressed in differential form:
Taylor expanding for small $\xi$ we obtain the identity%
\begin{equation}
\rho_{z,A+\partial\xi}=\rho_{zA}+\int d^{3}x\,\frac{\delta\rho_{zA}}{\delta
A_{ax}}\partial_{ax}\xi_{x}\approx\rho_{zA}~,
\end{equation}
which must hold for arbitrary choices of $\xi_{x}$. Therefore the allowed
$\rho$s are constrained to satisfy
\begin{equation}
\partial_{ax}\frac{\delta\rho_{zA}}{\delta A_{ax}}\approx0~. \label{GI b}%
\end{equation}
Analogous conditions apply to any gauge invariant functional.

The\ invariance of $\rho_{zA}$ must be preserved by time evolution and this
determines the gauge transformation of the drift potential $\varphi_{zA}$ and
the phase $\phi_{zA}$. From eqs.(\ref{rho dot a}), (\ref{curr velocity}), and
(\ref{phase}) we see that $\rho$ remains invariant under $A_{ax}\rightarrow
A_{ax}^{\xi}$ provided $\varphi_{zA}$ and $\phi_{zA}$ transform according to
the local gauge transformations,%
\begin{align}
\varphi_{zA}  &  \rightarrow\varphi_{zA}^{\xi}\approx\varphi_{zA}+%
{\displaystyle\sum\nolimits_{n}}
\frac{q_{n}}{c}\xi_{z_{n}}~,\label{GT c}\\
\phi_{zA}  &  \rightarrow\phi_{zA}^{\xi}\approx\phi_{zA}+%
{\displaystyle\sum\nolimits_{n}}
\frac{q_{n}}{c}\xi_{z_{n}}~, \label{GT d}%
\end{align}
where $\xi_{x}$ is a function on the $3d$-space $\mathcal{X}$. The dynamical
preservation of the gauge symmetry follows from the fact that the original
constraints (\ref{drift constraint}) and (\ref{gauge constraints}) are not
independent\ of each other --- both are linear in the same displacements
$\langle\Delta z_{n}^{a}\rangle$ --- which leads to the invariant combination
$(\partial_{na}\phi_{zA}-q_{n}A_{az_{n}})$ in (\ref{curr velocity}). Thus, our
particular choice of entropic prior and constraints has led to an ED\ that
allows a gauge-covariant HK flow.

We require that $\phi_{zA}$ transforms according to (\ref{GT d}), but being a
functional of $A$, we can also calculate $\phi_{zA}^{\xi}$ directly,%

\begin{equation}
\phi_{zA}^{\xi}=\phi_{zA^{\xi}}\quad\text{with}\quad\phi_{zA^{\xi}}=\phi
_{zA}+\int d^{3}x\,\frac{\delta\phi_{zA}}{\delta A_{ax}}\partial_{ax}\xi_{x}~.
\label{GT e}%
\end{equation}
Therefore, equating (\ref{GT e}) to (\ref{GT d}), the change in $\phi_{zA}$ is%
\begin{equation}
\delta_{\xi}\phi_{zA}=\int d^{3}x\,\frac{\delta\phi_{zA}}{\delta A_{ax}%
}\partial_{ax}\xi_{x}\approx%
{\displaystyle\sum\nolimits_{n}}
\frac{q_{n}}{c}\xi_{z_{n}}~,
\end{equation}
which, after integrating by parts and rearranging, gives
\begin{equation}
\int d^{3}x\,\left(  \partial_{ax}\frac{\delta\phi_{zA}}{\delta A_{ax}}+%
{\displaystyle\sum\nolimits_{n}}
\frac{q_{n}}{c}\delta_{z_{n}x}\right)  \xi_{x}\approx0~.
\end{equation}
Since this identity must hold for arbitrary choices of $\xi_{x}$ the allowed
phases are constrained to satisfy the infinite set of constraints (one at
every point $x\in\mathcal{X}$),
\begin{equation}
\partial_{ax}\frac{\delta\phi_{zA}}{\delta A_{ax}}+%
{\displaystyle\sum\nolimits_{n}}
\frac{q_{n}}{c}\delta_{z_{n}x}\approx0~. \label{GI c}%
\end{equation}

Having established the constraints on $\rho$s and $\phi$s we can find the
constraints on wave functions $\psi=\rho^{1/2}e^{i\phi/\hbar}$. Using
(\ref{GI a}) and (\ref{GT d}) and Taylor expanding gives
\begin{equation}
\psi_{zA}^{\xi}\approx\psi_{zA}\exp\left(  \frac{i}{\hbar}%
{\displaystyle\sum\nolimits_{n}}
\frac{q_{n}}{c}\xi_{z_{n}}\right)  =\psi_{zA}\left(  1+\frac{i}{\hbar}%
{\displaystyle\sum\nolimits_{n}}
\frac{q_{n}}{c}\xi_{z_{n}}\right)  ~. \label{GT f}%
\end{equation}
But we can also calculate $\psi_{zA}^{\xi}$ directly,%

\begin{equation}
\psi_{zA}^{\xi}=\psi_{_{z,A+\partial\xi}}=\psi_{zA}+\int d^{3}x\,\frac
{\delta\psi_{zA}}{\delta A_{ax}}\partial_{ax}\xi_{x}~.
\end{equation}
Therefore,
\begin{equation}
\delta_{\xi}\psi_{zA}=\int d^{3}x\,\frac{\delta\psi_{zA}}{\delta A_{ax}%
}\partial_{ax}\xi_{x}\approx\psi_{zA}\frac{i}{\hbar c}%
{\displaystyle\sum\nolimits_{n}}
q_{n}\xi_{z_{n}}~,
\end{equation}
which, after a little algebra, leads to
\begin{equation}
\left[  \int d^{3}x\,\xi_{x}\left(  \partial_{ax}\frac{\hbar c}{i}\frac
{\delta}{\delta A_{ax}}+%
{\displaystyle\sum\nolimits_{n}}
q_{n}\delta_{z_{n}x}\right)  \right]  \psi_{zA}\approx0~. \label{Gauss}%
\end{equation}
Once again, since $\xi_{x}$ is an arbitrary function, we conclude that the
requirement of gauge invariance restricts the allowed quantum states $\psi$ to
those that satisfy the infinite set of constraints (one at every point
$x\in\mathcal{X}$)
\begin{equation}
\left(  \partial_{ax}i\hbar c\frac{\delta}{\delta A_{ax}}-%
{\displaystyle\sum\nolimits_{n}}
q_{n}\delta_{z_{n}x}\right)  \psi_{zA}\approx0~, \label{Gauss a}%
\end{equation}
which is recognized as the quantum version of Gauss' law,
\begin{equation}
\left(  \partial_{ax}\hat{E}_{x}^{a}-\hat{\rho}_{x}^{\text{e}}\right)
\psi_{zA}\approx0~, \label{Gauss aa}%
\end{equation}
provided one identifies
\begin{equation}
\hat{E}^{a}(x)=i\hbar c\frac{\delta}{\delta A_{a}(x)} \label{electric field}%
\end{equation}
as the representation of the electric field operator $\hat{E}^{a}(x)$ acting
on functionals $\psi\lbrack z,A]$ of $A_{a}(x)$ and $\hat{\rho}_{x}^{\text{e}%
}$ is the electric charge density, eq.(\ref{densities}). In classical
electrodynamics eq.(\ref{electric field}) corresponds to the assertion that
the electric field (times $-1/c$) is the momentum that is canonically
conjugate to the potential $A_{a}(x)$.

Incidentally, the derivative operator $\hat{E}^{a}$ is how electric fields are
introduced in the ED approach. ED is founded on the recognition that the ontic
degrees of freedom are the radiation field which is redundantly represented by
the vector potential $A_{ax}$. And that is all there is; there is no such
thing as an \emph{ontic} electric field!

\subsection{Time evolution}

From now on we adopt the Hamiltonian $\hat{H}$ given by
eq.(\ref{Hamiltonian d}) and, to be specific,\ the potential $\hat{V}_{zA}$ is
chosen to be (\ref{potential}),
\begin{equation}
\hat{H}=\hat{H}^{\text{mat}}+\hat{H}^{\text{rad}}~,\label{Hamiltonian e}%
\end{equation}
where%
\begin{equation}
\hat{H}^{\text{mat}}=%
{\textstyle\sum\limits_{n}}
\frac{1}{2m_{n}}\delta^{ab}\frac{\hbar}{i}D_{na}\frac{\hbar}{i}D_{nb}%
\quad\text{and}\quad\hat{H}^{\text{rad}}=\frac{1}{2}\int d^{3}x\,(\hat{E}%
_{x}^{2}+\hat{B}_{x}^{2})~.\label{Hamiltonian ee}%
\end{equation}
The mere existence of the Gauss constraint, eq.(\ref{Gauss aa}), induces an
additional change. As pointed out by Dirac \cite{Dirac 1964}, it leads us to
consider the modified Hamiltonian,
\begin{equation}
\hat{H}_{\Phi}=\hat{H}+\hat{\Gamma}_{\Phi}\quad\text{with}\quad\hat{\Gamma
}_{\Phi}=\int d^{3}x\,\Phi_{xt}(\hat{\rho}_{x}^{\text{e}}-\partial_{ax}\hat
{E}_{x}^{a})\label{Ham Phi a}%
\end{equation}
where $\Phi_{xt}=\Phi_{xt}[z_{n},A]$ is some well-behaved but otherwise
arbitrary function of $x$, $t$, and of the ontic variables $z_{n}$ and
$A_{x}^{a}$.

\noindent\textbf{Remark:} In the Dirac approach to the \emph{classical}
dynamics of constrained Hamiltonian systems \cite{Dirac 1964} there is one
quantity $\Phi$ for each constraint, which accounts for $\Phi=\Phi_{xt}$ being
a function of $x$, and at each $x$ the quantity $\Phi_{xt}$ is an arbitrary
functional of the canonical variables $z_{n}$ and $A_{x}^{a}$ and \emph{their}
respective momenta. In the ED\ approach to \emph{quantum} dynamics there is an
important difference: the canonical variables are $\psi$ and $\psi^{\ast}$,
not $z_{n}$, $A_{x}^{a}$ and their momenta. The HK\ flows require that the
Hamiltonians $\tilde{H}$ and $\tilde{H}_{\Phi}$ be bilinear in $\psi$ and
$\psi^{\ast}$, eq.(\ref{Hamiltonian c}), which in turn implies that the
kernels $\hat{H}$ and $\hat{H}_{\Phi}$ must be \emph{independent} of the
canonical variables $\psi$ and $\psi^{\ast}$. Thus $\hat{\Gamma}_{\Phi}$ can
depend on $z_{n}$ and $A_{x}^{a}$ but not on $\psi$ and $\psi^{\ast}$.

The new Schr\"{o}dinger equation is
\begin{equation}
i\hbar\frac{\partial\psi}{\partial t}=\hat{H}_{\Phi}\psi=(\hat{H}+\hat{\Gamma
}_{\Phi})\psi~. \label{Ham Phi b}%
\end{equation}
Imposing $\hat{\Gamma}_{\Phi}\psi\approx0$ shows that the arbitrary function
$\Phi_{x}$ has no effect on the evolution of $\psi$, but it does introduce
indeterminism in the evolution of gauge dependent fields such as $\tilde
{A}_{ax}$ (see eq.(\ref{Adot b})).

The consistency of the whole scheme requires that the Gauss constraint be
preserved by time evolution, that is,%
\begin{equation}
\text{if\quad}\hat{\Gamma}_{\Phi}\psi\approx0\text{\quad then\quad}%
\partial_{t}(\hat{\Gamma}_{\Phi}\psi)\approx0~.\label{Gauss b}%
\end{equation}
More explicitly,
\begin{align*}
\partial_{t}(\hat{\Gamma}_{\Phi}\psi) &  =\int d^{3}x\,\dot{\Phi}_{xt}%
(\hat{\rho}_{x}^{\text{e}}-\partial_{ax}\hat{E}_{x}^{a})\psi+\frac{1}{i\hbar
}\int d^{3}x\,\Phi_{xt}(\hat{\rho}_{x}^{\text{e}}-\partial_{ax}\hat{E}_{x}%
^{a})\hat{H}_{\Phi}\psi\\
&  \approx\frac{1}{i\hbar}\int d^{3}x\,\Phi_{xt}\left(  \hat{H}(\hat{\rho}%
_{x}^{\text{e}}-\partial_{ax}\hat{E}_{x}^{a})\psi+[\hat{\rho}_{x}^{\text{e}%
}-\partial_{ax}\hat{E}_{x}^{a},\hat{H}]\psi\right)  ~,
\end{align*}
so that
\begin{equation}
\partial_{t}(\hat{\Gamma}_{\Phi}\psi)\approx\frac{1}{i\hbar}\int d^{3}%
x\,\Phi_{xt}[(\hat{\rho}_{x}^{\text{e}}-\partial_{ax}\hat{E}_{x}^{a}),\hat
{H}]\psi~,\label{Gauss bb}%
\end{equation}
which requires that we evaluate the commutator.

First we consider the commutators $[\hat{\rho}_{x}^{\text{e}},\hat
{H}^{\text{rad}}]$ and $[\partial_{ax}\hat{E}_{x}^{a},\hat{H}^{\text{rad}}]$.
The former vanishes trivially. For the latter we need $[\hat{E}_{x}^{a}%
,\hat{B}_{x^{\prime}}^{b}]$, which is%

\begin{equation}
\lbrack\hat{E}_{x}^{a},\hat{B}_{x^{\prime}}^{b}]\psi_{zA}=i\hbar
c\varepsilon^{bcd}\partial_{cx^{\prime}}[\frac{\delta}{\delta A_{ax}%
},A_{dx^{\prime}}]\psi_{zA}=i\hbar c\varepsilon^{abc}(\partial_{cx^{\prime}%
}\delta_{x^{\prime}x})\psi_{zA}~, \label{Comm EB}%
\end{equation}
and implies \quad\quad%
\begin{equation}
\lbrack\hat{E}_{x}^{a},\hat{H}^{\text{rad}}]\psi_{zA}=\frac{1}{2}\int
d^{3}x\,[\hat{E}_{x}^{a},\hat{B}_{x}^{2}]\psi_{zA}=i\hbar c(\partial_{x}%
\times\hat{B}_{x})^{a}\psi_{zA}~, \label{Comm EHrad}%
\end{equation}
and \quad\quad%
\begin{equation}
\lbrack\partial_{ax}\hat{E}_{x}^{a},\hat{H}^{\text{rad}}]\psi_{zA}=i\hbar
c\partial_{ax}(\partial_{x}\times\hat{B}_{x})^{a}\psi_{zA}=0~,
\label{Comm DEHrad}%
\end{equation}
Therefore,
\begin{equation}
\lbrack\hat{\rho}_{x}^{\text{e}}-\partial_{ax}\hat{E}_{x}^{a},\hat
{H}^{\text{rad}}]\psi_{zA}=0~. \label{Comm Gauss Hrad}%
\end{equation}

Next we deal with $[\partial_{ax}\hat{E}_{x}^{a},\hat{H}^{\text{mat}}]$ and
$[\hat{\rho}_{x}^{\text{e}},\hat{H}^{\text{mat}}]$. For the former use%

\begin{equation}
\lbrack\hat{E}_{x}^{a},D_{nb}]\psi_{zA}=q_{n}\delta_{b}^{a}\delta_{xz_{n}}%
\psi_{zA}~, \label{Comm ED}%
\end{equation}
and%
\begin{equation}
\delta^{bc}[\hat{E}_{x}^{a},D_{nb}D_{nc}]\psi_{zA}=2q_{n}\delta^{ab}%
\delta_{xz_{n}}D_{nb}\psi_{zA}+q_{n}\delta^{ab}\frac{\partial\delta_{xz_{n}}%
}{\partial z_{n}^{b}}\psi_{zA}~, \label{Comm EDD}%
\end{equation}
which, substituting $\hat{H}^{\text{mat}}$ from (\ref{Hamiltonian ee}), implies%

\begin{equation}
\partial_{ax}[\hat{E}_{x}^{a},\hat{H}^{\text{mat}}]\psi_{zA}=%
{\textstyle\sum\limits_{n}}
\frac{-\hbar^{2}q_{n}}{m_{n}}\left(  \delta^{ab}\partial_{ax}\delta_{xz_{n}%
}D_{nb}-\frac{1}{2}\partial_{x}^{2}\delta_{xz_{n}}\right)  \psi_{zA}~.
\label{Comm DE Hmat}%
\end{equation}
Next consider $[\hat{\rho}_{x}^{\text{e}},\hat{H}^{\text{mat}}]$. Use
(\ref{densities}) and (\ref{cov deriv}), so that%
\begin{equation}
\lbrack\hat{\rho}_{x}^{\text{e}},D_{na}]\psi=%
{\displaystyle\sum\nolimits_{n^{\prime}}}
q_{n^{\prime}}[\delta_{z_{n^{\prime}}x},\frac{\partial}{\partial z_{n}^{a}%
}-\frac{iq_{n}}{\hbar c}A_{az_{n}}]\psi=q_{n}\frac{\partial\delta_{z_{n}x}%
}{\partial x^{a}}\psi~,
\end{equation}
and%
\begin{align}
\delta^{ab}[\hat{\rho}_{x}^{\text{e}},D_{na}D_{nb}]\psi &  =\delta^{ab}%
[\hat{\rho}_{x}^{\text{e}},D_{na}]D_{nb}\psi+\delta^{ab}D_{na}[\hat{\rho}%
_{x}^{\text{e}},D_{nb}]\psi\\
&  =2q_{n}\delta^{ab}\frac{\partial\delta_{z_{n}x}}{\partial x^{a}}D_{nb}%
\psi-q_{n}\partial_{x}^{2}\delta_{xz_{n}}\psi~.
\end{align}
Therefore, substituting $\hat{H}^{\text{mat}}$ from (\ref{Hamiltonian ee}),
\begin{equation}
\lbrack\hat{\rho}_{x}^{\text{e}},\hat{H}^{\text{mat}}]\psi=%
{\textstyle\sum\limits_{n}}
\frac{-\hbar^{2}q_{n}}{m_{n}}\left(  \delta^{ab}\frac{\partial\delta_{z_{n}x}%
}{\partial z_{n}^{a}}D_{nb}-\frac{1}{2}\partial_{x}^{2}\delta_{xz_{n}}\right)
\psi~. \label{Comm rho Hmat}%
\end{equation}
Combining (\ref{Comm DE Hmat}) and (\ref{Comm rho Hmat}) leads to
\begin{equation}
\lbrack\hat{\rho}_{x}^{\text{e}}-\partial_{ax}\hat{E}_{x}^{a},\hat
{H}^{\text{mat}}]\psi_{zA}=0~, \label{Comm Gauss Hmat}%
\end{equation}
which, together with (\ref{Comm Gauss Hrad}), implies
\begin{equation}
\lbrack(\hat{\rho}_{x}^{\text{e}}-\partial_{ax}\hat{E}_{x}^{a}),\hat{H}]=0~,
\label{Gauss c}%
\end{equation}
and proves (\ref{Gauss bb}). Note that (\ref{Gauss c}) is not a weak equality;
it holds strongly.

We can also check the mutual consistency of time evolution and time-dependent
gauge transformations write (\ref{GT f}) as $\psi^{\xi}=\hat{U}_{\xi}\psi$
where $\hat{U}_{\xi}$ is the unitary operator
\begin{equation}
\hat{U}_{\xi}=\exp\left(  \frac{i}{\hbar c}%
{\displaystyle\sum\nolimits_{n}}
q_{n}\xi_{z_{n}}\right)  =\exp\left(  \frac{i}{\hbar c}\int d^{3}x\,\xi
_{x}\hat{\rho}_{x}^{\text{e}}\right)  ~.\label{U csi}%
\end{equation}
Then, the consistency of (\ref{Ham Phi b}) with the transformed
Schr\"{o}dinger equation,%
\begin{equation}
i\hbar\frac{\partial\psi^{\xi}}{\partial t}=\hat{H}_{\Phi}^{\xi}\psi^{\xi}~,
\end{equation}
requires that
\begin{equation}
i\hbar\frac{\partial}{\partial t}(\hat{U}_{\xi}\psi)=i\hbar\frac{\partial
\hat{U}_{\xi}}{\partial t}\psi+\hat{U}_{\xi}H_{\Phi}\psi=\hat{H}_{\Phi}^{\xi
}\hat{U}_{\xi}\psi
\end{equation}
or,%
\begin{equation}
\hat{H}_{\Phi}^{\xi}=\hat{U}_{\xi}H_{\Phi}\hat{U}_{\xi}^{-1}+i\hbar
\frac{\partial\hat{U}_{\xi}}{\partial t}\hat{U}_{\xi}^{-1}~,
\end{equation}
Substituting (\ref{U csi}), we find
\begin{equation}
\hat{H}_{\Phi}^{\xi}=\hat{H}_{\Phi}+\delta_{\xi}\hat{V}\quad\text{with}%
\quad\delta_{\xi}\hat{V}=-\frac{1}{c}\int d^{3}x\,\partial_{t}\xi_{x}\hat
{\rho}_{x}^{\text{e}}~.
\end{equation}
Thus, we recover the familiar fact that the Hamiltonian is not invariant under
time-dependent gauge transformations, and that the effect of the latter is to
cause a shift of the background scalar potential.

\subsection{Gauge transformations as HK flows}

Consider the Hamiltonian functional that describes the expected electric
charge density smeared by an arbitrary function $\zeta_{x}$,
\begin{align}
\tilde{Q}_{\zeta}^{\text{e}}  & =\int d^{3}x\,\zeta_{x}\left(  \int
dzDA\,\rho_{zA}\hat{\rho}_{x}^{\text{e}}\right)  \\
& =\int dzDA\,\psi_{zA}^{\ast}\hat{Q}_{\zeta}^{\text{e}}\psi_{zA}%
\quad\text{with}\quad\hat{Q}_{\zeta}^{\text{e}}=\int d^{3}x\,\zeta_{x}%
\hat{\rho}_{x}^{\text{e}}~.
\end{align}
It generates an HK flow $\psi_{zA}(\lambda)$ parametrized by $\lambda$ and
defined by the Hamilton/Schr\"{o}dinger equation, \noindent%
\begin{equation}
\frac{\partial\psi_{zA}}{\partial\lambda}=\{\psi_{zA},\tilde{Q}_{\zeta
}^{\text{e}}\}=\frac{1}{i\hbar}\frac{\delta\tilde{Q}_{\zeta}^{\text{e}}%
}{\delta\psi_{zA}^{\ast}}=\frac{1}{i\hbar}\hat{Q}_{\zeta}^{\text{e}}\psi
_{zA}~.
\end{equation}
The corresponding integral curve is
\begin{equation}
\psi_{zA}(\lambda)=\psi_{zA}(0)\exp\left(  \frac{\lambda}{i\hbar}\hat
{Q}_{\zeta}^{\text{e}}\right)  =\psi_{zA}(0)\exp\left(  \frac{\lambda}{i\hbar
}\int d^{3}x\,\zeta_{x}\hat{\rho}_{x}^{\text{e}}\right)  ~.
\end{equation}
Comparing with (\ref{GT f}) this is recognized as the gauge orbit generated by
gauge transformations with function $\xi_{x}=-c\lambda\zeta_{x}$.\ In other
words, the smeared expected charge density $\tilde{Q}_{\zeta}^{\text{e}}$ is
the generator of gauge transformations.

We can also consider the Hamiltonian function that describes the smeared Gauss
constraint, eq.(\ref{Ham Phi a}),%
\begin{equation}
\tilde{\Gamma}_{\Phi}=\int dzDA\,\psi_{zA}^{\ast}\hat{\Gamma}_{\Phi}\psi
_{zA}\quad\text{with}\quad\hat{\Gamma}_{\Phi}=\int d^{3}x\,\Phi_{xt}(\hat
{\rho}_{x}^{\text{e}}-\partial_{ax}\hat{E}_{x}^{a})~.
\end{equation}
The corresponding HK\ flow $\psi_{zA}(\lambda)$ is given by
\begin{equation}
\frac{\partial\psi_{zA}}{\partial\lambda}=\{\psi_{zA},\tilde{\Gamma}_{\Phi
}\}=\frac{1}{i\hbar}\frac{\delta\tilde{\Gamma}_{\Phi}}{\delta\psi_{zA}^{\ast}%
}=\frac{1}{i\hbar}\tilde{\Gamma}_{\Phi}\psi_{zA}~
\end{equation}
and its integral curves are
\begin{equation}
\psi_{zA}(\lambda)=\exp\left(  \frac{\lambda}{i\hbar}\tilde{\Gamma}_{\Phi
}\right)  \psi_{zA}(0)~.\label{Gauss int curve}%
\end{equation}
We see that if $\hat{\Gamma}_{\Phi}\psi_{zA}(0)\approx0$ then $\hat{\Gamma
}_{\Phi}\psi_{zA}(\lambda)\approx0$ too, that is, $\tilde{\Gamma}_{\Phi}$
generate curves along which the Gauss constraint is obeyed.

\noindent\textbf{Remark}: According to Dirac's hypothesis constraints that
commute with all other constraints (called first-class constraints) are the
generators of gauge transformations [Dirac 1964]. Even though Dirac's
hypothesis holds true for those cases he tested, in the context of ED the
Dirac hypothesis fails. Indeed, we can easily check that the commutators
\begin{equation}
\lbrack\hat{\Gamma}_{\Phi},\hat{\Gamma}_{\Phi^{\prime}}]=0~
\end{equation}
vanish strongly so that the constraints $\hat{\Gamma}_{\Phi}$ are first class,
but eq.(\ref{Gauss int curve}) clearly shows that $\hat{\Gamma}_{\Phi}$ does
not generate gauge transformations.

\subsection{Maxwell equations}

\noindent We have just concluded our formulation of the ED\ approach to
non-relativistic quantum electrodynamics. In the next two sections we shall
offer some further evidence to the effect that despite ED's unorthodox
foundations which include a clarity about which variables play an ontic and
which an epistemic role and the absence of an ontic dynamics, the ED model
developed here is empirically equivalent to the standard versions of QED.

We saw, back in eqs.(\ref{potential}) and (\ref{electric field}), that the
electric $E_{x}^{a}$ and the magnetic $B_{x}^{a}$ fields are not fundamental;
they are derivative, auxiliary concepts related to the concept that is
actually fundamental, which is the vector potential $A_{x}^{a}$. Let
\begin{align}
\tilde{A}_{x}^{a} &  =\int dzDA\,\psi_{zA}^{\ast}\hat{A}_{x}^{a}\psi_{zA}%
\quad\text{with}\quad\hat{A}_{x}^{a}\overset{\text{def}}{=}A_{x}^{a}~,\\
\tilde{B}_{x}^{a} &  =\int dzDA\,\psi_{zA}^{\ast}\hat{B}_{x}^{a}\psi_{zA}%
\quad\text{with}\quad\hat{B}_{x}^{a}\overset{\text{def}}{=}(\partial\times
\hat{A})_{x}^{a}~,\\
\tilde{E}_{x}^{a} &  =\int dzDA\,\psi_{zA}^{\ast}\hat{E}_{x}^{a}\psi_{zA}%
\quad\text{with}\quad\hat{E}_{x}^{a}\overset{\text{def}}{=}i\hbar
c\frac{\delta}{\delta A_{ax}}~.
\end{align}
Two of the Maxwell equations we have already met. The absence of magnetic
monopoles follows from the definition of the $\hat{B}_{x}$ field,
eq.(\ref{potential}),%
\begin{equation}
\hat{B}_{x}\overset{\text{def}}{=}\partial_{x}\times\hat{A}_{x}\Rightarrow
\tilde{B}=\partial_{x}\times\tilde{A}_{x}\Rightarrow\partial_{x}\cdot\tilde
{B}_{x}=0~.
\end{equation}
Gauss' law is the Gauss constraint, eq.(\ref{Gauss aa}), to be imposed on wave
functions,
\begin{equation}
\left(  \partial_{ax}\hat{E}_{x}^{a}-\hat{\rho}_{x}^{\text{e}}\right)
\psi_{zA}\approx0\quad\text{or}\quad\partial_{ax}\tilde{E}_{x}^{a}%
\approx\tilde{\rho}_{x}^{\text{e}}~.
\end{equation}
The other two Maxwell equations are equations of time evolution for $\tilde
{A}_{x}^{a}$, $\tilde{B}_{x}^{a}$, and $\tilde{E}_{x}^{a}$,
\begin{equation}
\partial_{t}\tilde{A}_{x}^{a}=\{\tilde{A}_{x}^{a},\tilde{H}_{\Phi}%
\}\quad\text{and}\quad\partial_{t}\tilde{E}_{x}=\{\tilde{E}_{x}^{a},\tilde
{H}_{\Phi}\}~.
\end{equation}

Use (\ref{Hamiltonian ee}) and (\ref{Ham Phi a}) to write the modified
Hamiltonian as
\begin{equation}
\hat{H}_{\Phi}=\hat{H}_{\Phi}^{\text{mat}}+\hat{H}_{\Phi}^{\text{rad}%
}~,\label{Hamiltonian f}%
\end{equation}
where%
\begin{equation}
\hat{H}_{\Phi}^{\text{mat}}=\hat{H}^{\text{mat}}+\int d^{3}x\,\Phi_{x}%
\hat{\rho}_{x}^{\text{e}}\quad\text{and}\quad\hat{H}_{\Phi}^{\text{rad}}%
=\hat{H}^{\text{rad}}+\int d^{3}x\,\hat{E}_{x}^{a}\partial_{ax}\Phi
_{x}\label{Hamiltonian ff}%
\end{equation}
where the last term has been integrated by parts. Calculating the PBs leads
to
\begin{align}
\partial_{t}\tilde{A}_{x}^{a} &  =\frac{1}{i\hbar}\int dzDA\,\psi_{zA}^{\ast
}[A_{x}^{a},\hat{H}_{\Phi}]\psi_{zA}~,\label{Adot a}\\
\partial_{t}\tilde{E}_{x}^{a} &  =\frac{1}{i\hbar}\int dzDA\,\psi_{zA}^{\ast
}[\hat{E}_{x}^{a},\hat{H}_{\Phi}]\psi_{zA}~.\label{Edot}%
\end{align}
The $A$-commutators give
\begin{equation}
\lbrack A_{x}^{a},\hat{H}_{\Phi}^{\text{mat}}]\psi_{zA}=0\quad\text{and}%
\quad\lbrack A_{x}^{a},\hat{H}_{\Phi}^{\text{rad}}]\psi_{zA}=-i\hbar
c\,\left(  \hat{E}_{x}^{a}+\partial_{x}^{a}\Phi_{x}\right)  \psi
_{zA}~.\label{Comm AH}%
\end{equation}
The first equation is trivial; the second only slightly less so. Use
\begin{equation}
\lbrack A_{x}^{a},\hat{E}_{x^{\prime}}^{b}]\psi=i\hbar c[A_{x}^{a}%
,\frac{\delta}{\delta A_{bx^{\prime}}}]\psi=-i\hbar c\delta^{ab}%
\delta_{xx^{\prime}}~,
\end{equation}
and (\ref{Comm AH}) follows.

Substituting into (\ref{Adot a}), the time evolution of $\tilde{A}$ is
\begin{equation}
\partial_{t}\tilde{A}_{x}^{a}=-c(\tilde{E}_{x}^{a}+\partial_{x}^{a}\Phi
_{x})~.\label{Adot b}%
\end{equation}
This shows that the longitudinal component of $\partial_{t}\tilde{A}^{a}$ is
quite arbitrary, which is the defining characteristic of gauge fields. We can
make this explicit by decomposing $A_{x}^{a}$ into its longitudinal and
transverse parts,
\begin{equation}
A_{x}^{a}=A_{\ell x}^{a}+A_{\text{tr}x}^{a}~,
\end{equation}
with
\begin{equation}
\partial_{x}\times A_{\ell x}=0\quad\text{and}\quad\partial_{x}\cdot
A_{\text{tr}x}=0~,
\end{equation}
so that $A_{\ell}^{a}$ is the gauge dependent component and $A_{\text{tr}}%
^{a}$ is gauge invariant. We can also write (\ref{Adot b})
\begin{equation}
\tilde{E}_{x}^{a}=-\partial_{x}^{a}\Phi_{x}-\frac{1}{c}\partial_{t}\tilde
{A}_{x}^{a}\label{E pots}%
\end{equation}
which, if we interpret the function $\Phi_{xt}$ as a scalar potential,
recovers the familiar expression for the electric field in terms of the
potentials $\Phi_{x}$ and $\tilde{A}_{x}^{a}$. At this point we can
\emph{extend} the gauge transformations to the scalar potential $\Phi_{xt}$,
and write
\begin{equation}
\tilde{A}_{\text{tr}x}^{\xi}=\tilde{A}_{\text{tr}x}~,\quad\tilde{A}_{\ell
x}^{\xi}=\tilde{A}_{\ell x}+\partial_{x}\xi_{x}\quad\text{and}\quad\Phi
_{x}^{\xi}=\Phi_{x}-\frac{1}{c}\xi_{x}~,\label{GT g}%
\end{equation}
which makes the electric field (\ref{E pots}) gauge invariant. This is a shift
of perspective that deserves to be emphasized: In standard approaches to
electrodynamics, whether classical or quantum, an identity such as
(\ref{E pots}) is taken as part of the definition of the potentials from the
already given, presumably fundamental, field strengths. In the ED approach
(\ref{E pots}) is a derived result. The fundamental equation is (\ref{Adot b}%
); it describes the time evolution of the field $\tilde{A}_{x}^{a}$, in terms
of quantities, $\tilde{E}_{x}^{a}$ and $\Phi_{x}$, introduced by the Gauss
constraint, eqs.(\ref{Gauss aa})-(\ref{Ham Phi a}). The third Maxwell equation
follows immediately; just take the curl of (\ref{E pots}) to find Faraday's
law,
\begin{equation}
\partial_{x}\times\tilde{E}_{x}=-\frac{1}{c}\partial_{t}\tilde{B}%
_{x}~.\label{Faraday}%
\end{equation}

Next we deal with (\ref{Edot}). Since
\begin{align}
&  [\hat{E}_{x}^{a},\hat{\Gamma}_{\Phi}]\psi=i\hbar c\int d^{3}x\,[\frac
{\delta}{\delta A_{ax}},\,\Phi_{x}(\hat{\rho}_{x}^{\text{e}}-\partial_{ax}%
\hat{E}_{x}^{a})]\psi\nonumber\\
&  =i\hbar c\int d^{3}x\,\left(  \Phi_{x}[\frac{\delta}{\delta A_{ax}}%
,\,(\hat{\rho}_{x}^{\text{e}}-\partial_{ax}\hat{E}_{x}^{a})]+[\frac{\delta
}{\delta A_{ax}},\,\Phi_{x}](\hat{\rho}_{x}^{\text{e}}-\partial_{ax}\hat
{E}_{x}^{a})\right)  \psi\approx0~,
\end{align}
we can replace $(\hat{H}_{\Phi}^{\text{rad}},\hat{H}_{\Phi}^{\text{mat}})$ by
$(\hat{H}^{\text{rad}},\hat{H}^{\text{mat}})$ in the commutators. The
commutator $[\hat{E}_{x}^{a},\hat{H}^{\text{rad}}]$ is given in
(\ref{Comm EHrad}). The commutator $[\hat{E}_{x}^{a},\hat{H}^{\text{mat}}]$
requires more algebra; use (\ref{Comm EDD}) and substitute into%

\begin{equation}
\frac{1}{i\hbar}\int dzDA\,\psi^{\ast}[\hat{E}_{x}^{a},\hat{H}^{\text{mat}%
}]\psi=\int dzDA\,%
{\textstyle\sum\limits_{n}}
\frac{i\hbar q_{n}}{m_{n}}\delta^{ab}\delta_{xz_{n}}(\psi^{\ast}D_{nb}%
\psi-\frac{1}{2}\frac{\partial\rho_{zA}}{\partial z_{n}^{b}})~.
\end{equation}
This can be simplified considerably: From $\psi=\rho^{1/2}e^{i\phi/\hbar}$, we
obtain the identity,
\begin{equation}
\psi_{zA}^{\ast}\frac{\partial}{\partial z_{n}^{b}}\psi_{zA}-\frac{1}{2}%
\frac{\partial\rho_{zA}}{\partial z_{n}^{b}}=\frac{i}{\hbar}\psi_{zA}^{\ast
}\frac{\partial\phi_{zA}}{\partial z_{n}^{b}}\psi_{zA}~,
\end{equation}
which when combined with eqs.(\ref{curr velocity}) and (\ref{dens current})
leads to%
\begin{equation}
\frac{1}{i\hbar}\int dzDA\,\psi_{zA}^{\ast}[\hat{E}_{x}^{a},\hat
{H}^{\text{mat}}]\psi_{zA}=\int dzDA\,\rho_{zA}%
{\displaystyle\sum\nolimits_{n}}
q_{n}\delta_{z_{n}x}v_{n}^{a}=-\tilde{J}_{x}^{\text{e}a}~. \label{Comm EHmatt}%
\end{equation}
Finally, substitute (\ref{Comm EHrad}) and (\ref{Comm EHmatt}) into
(\ref{Edot}) to find the fourth Maxwell equation, the Amp\`{e}re-Maxwell law,%
\begin{equation}
\partial_{t}\tilde{E}_{x}=c\partial_{x}\times\tilde{B}_{x}-\tilde{J}%
_{x}^{\text{e}}~. \label{Amp-Max}%
\end{equation}

This derivation of the Maxwell equations has taken full account of the
redundancy of the representation of the $A_{x}^{a}$ field but it has been
carried out without having specified a gauge condition. In other words, the
derivation holds for all gauges. However, in applications there is no reason
not take advantage of the freedom to choose a convenient gauge; this we do next.

\subsection{The Coulomb potential}

The Gauss constraint is a restriction on the longitudinal component of the
electric field operator. This suggests decomposing the vector operator
$\hat{E}_{x}$ into its longitudinal and transverse parts,
\begin{equation}
\hat{E}=\hat{E}_{\ell}+\hat{E}_{\text{tr}}\quad\text{with}\quad\partial
_{x}\times\hat{E}_{\ell x}=0\quad\text{and}\quad\partial_{x}\cdot\hat
{E}_{\text{tr}x}=0 \label{LT parts a}%
\end{equation}
where, using (\ref{E pots}),
\begin{equation}
\tilde{E}_{\ell x}^{a}=-\partial_{x}^{a}\Phi_{x}-\frac{1}{c}\partial_{t}%
\tilde{A}_{\ell x}^{a}~.
\end{equation}
We can take advantage of (\ref{GT g}) to impose the Coulomb gauge,
\begin{equation}
\tilde{A}_{\ell x}^{a}=0\quad\text{or}\quad\partial_{a}\tilde{A}_{x}^{a}=0~.
\end{equation}
Then $\hat{E}_{\ell\,}$ can be written as the gradient of a scalar potential
operator $\hat{\Phi}_{x}^{C}$,%
\begin{equation}
\hat{E}_{\ell x}=-\partial_{x}\hat{\Phi}_{x}^{C}~,\quad\hat{E}_{\text{tr}%
x}=\hat{E}_{x}+\partial_{x}\hat{\Phi}_{x}^{C}~, \label{LT parts c}%
\end{equation}
and the Gauss constraint reads
\begin{equation}
\left(  \partial_{a}\hat{E}_{\ell}^{a}-\hat{\rho}^{\text{e}}\right)
\psi\approx0\quad\text{or}\quad\left(  \partial_{a}\partial^{a}\hat{\Phi}%
^{C}+\hat{\rho}^{\text{e}}\right)  \psi\approx0~, \label{Gauss f}%
\end{equation}
which is the quantum version of the Poisson equation. The familiar solution of
(\ref{Gauss f}) is the Coulomb potential,
\begin{equation}
\hat{\Phi}_{x}^{C}\psi\approx\int d^{3}x^{\prime}\frac{\hat{\rho}_{x^{\prime}%
}^{\text{e}}}{4\pi|\vec{x}-\vec{x}^{\prime}|}\psi=\sum_{n}\frac{q_{n}}%
{4\pi|\vec{x}-\vec{z}_{n}|}\psi~,
\end{equation}
where we used (\ref{densities}). This reproduces a well-known result in QED,
namely, that the scalar potential $\hat{\Phi}_{x}^{C}(z)$ is not an
independent dynamical variable but a function of the particle positions. (Note
that $\Phi_{x}^{C}$ is not a function of the canonical variables $\psi$ and
$\psi^{\ast}$.)

In terms of $\hat{E}_{\ell x}$ and $\hat{E}_{\text{tr}x}$, and imposing the
Gauss constraint, the Hamiltonian (\ref{Hamiltonian f}) now reads
\begin{equation}
\hat{H}\approx%
{\textstyle\sum\limits_{n}}
\frac{-\hbar^{2}}{2m_{n}}\delta^{ab}D_{na}D_{nb}+\int d^{3}x\,\frac{1}%
{2}\delta_{ab}\left(  \partial^{a}\hat{\Phi}^{C}\partial^{b}\hat{\Phi}%
^{C}+\hat{E}_{\text{tr}}^{a}\hat{E}_{\text{tr}}^{b}+\hat{B}^{a}\hat{B}%
^{b}\right)  ~.
\end{equation}
Integrating the Coulomb term by parts and using the Poisson equation yields
the familiar result
\begin{equation}
\int d^{3}x\,\frac{1}{2}\delta_{ab}\partial^{a}\hat{\Phi}^{C}\partial^{b}%
\hat{\Phi}^{C}=\frac{1}{2}\int d^{3}x\,\hat{\Phi}^{C}\hat{\rho}^{\text{e}%
}=\frac{1}{2}\sum_{n\neq n^{\prime}}\frac{q_{n}q_{n^{\prime}}}{4\pi|\vec
{z}_{n}-\vec{z}_{n^{\prime}}|}~,
\end{equation}
where we have invoked the usual \textquotedblleft argument\textquotedblright%
\ that the infinite constant contributed by the self-energy terms
$n=n^{\prime}$ can be dropped because it has no effect on energy differences
and transition rates. (Clearly the present non-relativistic formulation is not
adequate to deal with the self-energy terms.) We can shift the Coulomb term
from $\hat{H}^{\text{rad}}$ to $\hat{H}^{\text{matt}}$ and write $\hat{H}$ in
the standard form,%
\begin{equation}
\hat{H}=\hat{H}^{\text{rad}}+\hat{H}^{\text{mat}}~,\label{Hqed}%
\end{equation}
with
\begin{equation}
\hat{H}^{\text{rad}}=\int d^{3}x\,\frac{1}{2}\delta_{ab}\left(  \hat
{E}_{\text{tr}}^{a}\hat{E}_{\text{tr}}^{b}+\hat{B}^{a}\hat{B}^{b}\right)
~,\label{Hrad}%
\end{equation}%
\begin{equation}
\hat{H}^{\text{mat}}=%
{\textstyle\sum\limits_{n}}
\frac{-\hbar^{2}}{2m_{n}}\delta^{ab}D_{na}D_{nb}+\frac{1}{2}\sum_{n\neq
n^{\prime}}\frac{q_{n}q_{n^{\prime}}}{4\pi|\vec{z}_{n}-\vec{z}_{n^{\prime}}%
|}~.\label{Hmat}%
\end{equation}
We have reached familiar QED territory. Even though the ED ontic/epistemic
interpretation differs from the standard textbook presentations, the ED
equations reproduce the standard formalism of QED in the Coulomb gauge.

\subsection{The action}

In the ED framework the introduction of an action principle is not
fundamental; it is a convenient way to summarize the content of an already
established formalism. The idea is to construct an action that reproduces the
equations of motion and the constraints.

Consider a region $R$ of the e-phase space and a time interval $T$ that
extends from \thinspace$t_{1}$ to $t_{2}$, and define the action functional
\begin{align}
\mathcal{A}[\Phi,\psi,\psi^{\ast}]  & =\int_{RT}dt\,dzDA\,\psi^{\ast}\left(
i\hbar\partial_{t}-\hat{H}_{\Phi}\right)  \psi\label{action a}\\
& =\int_{RT}dt\,dzDA\,\psi^{\ast}\left(  i\hbar\partial_{t}-\hat{H}\ -\int
d^{3}x\,\Phi_{xt}(\hat{\rho}_{x}^{\text{e}}-\partial_{ax}\hat{E}_{x}%
^{a})\right)  \psi\ .\nonumber
\end{align}
where $\tilde{H}$ is given in (\ref{Hamiltonian d}). The equations of motion
and the Gauss constraint are obtained by requiring that the action be
stationary under variations $\delta\psi$ and $\delta\psi^{\ast}$ that vanish
at the boundary of the region $RT$. The functional $\Phi_{xt}[zA]$ is a
Lagrange multiplier that enforces the Gauss constraint. Then
\begin{align}
\delta\mathcal{A}= &  \int_{TR}dt\,dzDA\,\left[  \delta\psi^{\ast}\left(
i\hbar\partial_{t}\psi-\hat{H}\psi-\hat{\Gamma}_{\Phi}\psi\right)  -\left(
i\hbar\partial_{t}\psi^{\ast}+\hat{H}^{\ast}\psi^{\ast}+\hat{\Gamma}_{\Phi
}^{\ast}\psi^{\ast}\right)  \delta\psi\right.  \nonumber\\
&  \left.  +~\psi^{\ast}\int d^{3}x\,\delta\Phi_{xt}\left(  \partial_{ax}%
\hat{E}_{x}^{a}-\hat{\rho}_{x}^{e}\right)  \psi\right]  =0.
\end{align}
The $\delta\Phi_{xt}[zA]$ variation leads to the constraint
\begin{equation}
\left(  \partial_{a}\hat{E}^{a}-\hat{\rho}^{e}\right)  \psi\approx
0\quad\text{or}\quad\hat{\Gamma}_{\Phi}\psi\approx0~,\label{Gauss e}%
\end{equation}
while $\delta\psi^{\ast}$ and $\delta\psi$ lead to
\begin{equation}
i\hbar\partial_{t}\psi=(\hat{H}\,+\hat{\Gamma}_{\Phi})\psi\label{Sch c}%
\end{equation}
and its complex conjugate.

\section{Final remarks}

With the derivation of the QED Hamiltonian, whether in an unspecified gauge,
eq.(\ref{Hamiltonian f}),\ or in the Coulomb gauge, eq.(\ref{Hqed}), the
ED\ formulation of QED reaches a certain level of completion. On the basis of
entropic inference we have not just derived an electrodynamics; we have
derived a \emph{quantum} electrodynamics. We have derived the linear
Schr\"{o}dinger equation with its attendant superposition principle. Along the
way we have justified the need for complex numbers and the arrow of time.
These are general consequences of the ED\ framework. Here we have expanded the
framework to include invariance under local gauge transformations. This has
allowed us to rederive charge conservation, the Maxwell equations, the Coulomb
potential, and the familiar QED equations that have historically received so
much empirical support. And, of course, the understanding of local gauge
invariance within the ED setting is a necessary stepping stone towards
relativistic quantum field theories such as Yang-Mills theories and gravity.

Much, however, remains to be done. Most glaring is the absence of any mention
of the concept of photon. We have two reasons not to pursue those applications
where the concept of photon becomes the indispensable tool. One reason is that
this paper is already too long; it has to end somewhere. Another reason is
that the actual study of the quantum correlations between particles and
fields, the transition rates of photon emission and absorption of light, and
so much more can be pursued by standard techniques. But there is one important
remark concerning photons that must be included here.

Photons are the names we give to those special quantum states --- the wave
functions of certain excited states of the radiation field\ --- that provide
the indispensable vocabulary to analyze exchanges of energy and momentum.
Therefore, even before one addresses the details of the construction of Fock
spaces for the radiation field, the ED framework allows the following possibly
unsettling conclusion: \emph{Photons are not particles and, even worse, they
are not real}. The point being made is simple. The ED philosophy insists in a
clear ontological commitment: Particles and radiation fields are real, they
belong to the ontic sector. Probabilities and wave functions are not real,
they belong to the epistemic sector. They are tools we have developed to
figure out what the real stuff might be doing --- what might the ontic
microstates be and how they might be changing. Perhaps surprisingly,
quantities such as energies, linear and angular momenta, electric currents and
even photons all belong to the epistemic sector. They are not properties of
the ontic particles or the ontic fields but concepts that are useful in the
analysis of the dynamics of probabilities.

\section{Appendix: The Continuity Equation}

The goal is to rewrite the integral equation for the evolution of probability,
eq.(\ref{DefTime}), in differential form as a continuity equation. We proceed
by adapting a technique \cite{Caticha 2025} borrowed from diffusion theory
\cite{Chandrasekhar 1943}. Multiply eq.(\ref{DefTime}) by a smooth test
functional $f_{z^{\prime}A^{\prime}}$ and integrate over $(z^{\prime
},A^{\prime})$ with measure $dz^{\prime}DA^{\prime}$,
\begin{equation}
\int dz^{\prime}DA^{\prime}\rho_{z^{\prime}A^{\prime}}^{\prime}f_{z^{\prime
}A^{\prime}}=\int dzDA\left[  \int dz^{\prime}DA^{\prime}\,P_{z^{\prime
}A^{\prime}|zA}f_{z^{\prime}A^{\prime}}\right]  \rho_{zA}~. \tag{A1}%
\end{equation}
The test functional $f_{z^{\prime}A^{\prime}}$ is assumed sufficiently smooth
precisely so that orders of integration can be swapped and we can Taylor
expand about $(z,A)$. Since $\Delta z$ and $\Delta A$ are of $O(\Delta
t^{1/2})$, keeping terms up to $O(\Delta t)$, then as $\Delta t\rightarrow0$
the integral in the brackets is%
\begin{align}
\left[  \cdots\right]   &  =\int dz^{\prime}DA^{\prime}\,P_{z^{\prime
}A^{\prime}|zA}\left(  f_{zA}+%
{\textstyle\sum\nolimits_{n}}
\frac{\partial f}{\partial z_{n}^{a}}\Delta z_{n}^{a}+\frac{1}{2}%
{\textstyle\sum\nolimits_{nn^{\prime}}}
\frac{\partial^{2}f}{\partial z_{n}^{a}\partial z_{n^{\prime}}^{b}}\Delta
z_{n}^{a}\Delta z_{n^{\prime}}^{b}\right. \nonumber\\
&  +\int d^{3}x\,\frac{\delta f}{\delta A_{ax}}\Delta A_{ax}+\frac{1}{2}\int
d^{3}x_{1}d^{3}x_{2}\,\frac{\delta^{2}f}{\delta A_{ax_{1}}\delta A_{bx_{2}}%
}\Delta A_{ax_{1}}\Delta A_{bx_{2}}+\nonumber\\
&  +%
{\textstyle\sum\nolimits_{n}}
\left.  \int d^{3}x\,\frac{\partial}{\partial z_{n}^{a}}(\frac{\delta
f}{\delta A_{bx}})\Delta z_{n}^{a}\Delta A_{bx}\right)  +o(\Delta t)~.
\tag{A2}%
\end{align}
Rearrange using eqs.(\ref{drift fluct a}) and (\ref{drift fluct b}),
\begin{align}
\left[  \cdots\right]   &  =f_{zA}+\Delta t%
{\textstyle\sum\nolimits_{n}}
\left(  \left[  \frac{\delta^{ab}}{m_{n}}\left(  \partial_{nb}\varphi
_{zA}-\beta_{n}A_{bz_{n}}\right)  \right]  \frac{\partial f}{\partial
z_{n}^{a}}+\frac{1}{2}\frac{\partial^{2}f}{\partial z_{n}^{a}\partial
z_{n}^{b}}\frac{\eta}{m_{n}}\delta^{ab}\right) \nonumber\\
&  +\Delta t\int d^{3}x\,c^{2}\delta_{ab}\left(  \frac{\delta\varphi_{zA}%
}{\delta A_{bx}}\frac{\delta f}{\delta A_{ax}}+\frac{\eta}{2}\frac{\delta
^{2}f}{\delta A_{ax}\delta A_{bx}}\right)  ~. \tag{A3}%
\end{align}
Substituting into the RHS of (A1),
\begin{align}
\text{RHS}  &  =\int dzDAf_{zA}\rho_{zA}\nonumber\\
&  +\Delta t\int dzDA%
{\textstyle\sum\nolimits_{n}}
\left(  \left[  \frac{\delta^{ab}}{m_{n}}\left(  \partial_{nb}\varphi
_{zA}-\beta_{n}A_{bz_{n}}\right)  \right]  \frac{\partial f}{\partial
z_{n}^{a}}+\frac{\eta}{2m_{n}}\delta^{ab}\frac{\partial^{2}f}{\partial
z_{n}^{a}\partial z_{n}^{b}}\right)  \rho_{zA}\nonumber\\
&  +\Delta t\int dzDA\int d^{3}x\,c^{2}\delta_{ab}\left(  \frac{\delta
\varphi_{zA}}{\delta A_{bx}}\frac{\delta f}{\delta A_{ax}}+\frac{\eta}{2}%
\frac{\delta^{2}f}{\delta A_{ax}\delta A_{bx}}\right)  \rho_{zA}~. \tag{A4}%
\end{align}
Next use,
\begin{align}
\int dz^{\prime}DA^{\prime}\rho_{z^{\prime}A^{\prime}}^{\prime}f_{z^{\prime
}A^{\prime}}-\int dzDA\rho_{zA}f_{zA}  &  =\int dzDA\left(  \rho_{zA}^{\prime
}-\rho_{zA}\right)  f_{zA}\nonumber\\
&  =\Delta t\int dzDA\frac{\partial\rho_{zA}}{\partial t}f_{zA}~, \tag{A5}%
\end{align}
integrate (A4) by parts, and divide by $\Delta t$ to get%
\begin{align}
\int dzDA\frac{\partial\rho_{zA}}{\partial t}f_{zA}  &  =\int dzDA%
{\textstyle\sum\nolimits_{n}}
\left(  \frac{\partial}{\partial z_{n}^{a}}\left[  -\rho_{zA}\frac{\delta
^{ab}}{m_{n}}\left(  \partial_{nb}\varphi_{zA}-\beta_{n}A_{bz_{n}}\right)
\right]  \right. \nonumber\\
&  +\left.  \frac{\eta}{2m_{n}}\delta^{ab}\frac{\partial^{2}\rho_{zA}%
}{\partial z_{n}^{a}\partial z_{n}^{b}}\right)  f_{zA}\nonumber\\
&  \int dzDA\int d^{3}x\,c^{2}\delta_{ab}\left(  -\frac{\delta}{\delta A_{ax}%
}\left[  \rho_{zA}\frac{\delta\varphi_{zA}}{\delta A_{bx}}\right]  +\frac
{\eta}{2}\frac{\delta^{2}\rho_{zA}}{\delta A_{ax}\delta A_{bx}}\right)
f_{zA}~. \tag{A6}%
\end{align}
This must hold for arbitrary test functionals $f_{zA}$. Therefore we get a
differential equation,
\begin{align}
\frac{\partial\rho_{zA}}{\partial t}  &  =-%
{\textstyle\sum\nolimits_{n}}
\frac{\partial}{\partial z_{n}^{a}}\left[  \rho_{zA}\left(  \frac{\delta^{ab}%
}{m_{n}}\left(  \partial_{nb}\varphi_{zA}-\beta_{n}A_{bz_{n}}\right)
-\frac{\eta}{2m_{n}}\delta^{ab}\frac{\partial\log\rho_{zA}}{\partial z_{n}%
^{b}}\right)  \right] \nonumber\\
&  -\int d^{3}x\,c^{2}\delta_{ab}\frac{\delta}{\delta A_{ax}}\left[  \rho
_{zA}\left(  \frac{\delta\varphi_{zA}}{\delta A_{bx}}-\frac{\eta}{2}%
\frac{\delta\log\rho_{zA}}{\delta A_{bx}}\right)  \right]  ~, \tag{A7}%
\end{align}
which is a continuity equation for the probability flow $\rho_{zA}$ in the
ontic configuration space
\begin{equation}
\frac{\partial\rho_{zA}}{\partial t}=-%
{\textstyle\sum\nolimits_{n}}
\frac{\partial}{\partial z_{n}^{a}}\left(  v_{n}^{a}\rho_{zA}\right)  -\int
d^{3}x\,\frac{\delta}{\delta A_{ax}}\left(  u_{ax}\rho_{zA}\right)  \tag{A8}%
\end{equation}
with a current velocity $(v,u)$ that has components along $z_{n}^{a}$,
\begin{equation}
v_{n}^{a}[z,A]=\frac{\delta^{ab}}{m_{n}}\left(  \partial_{nb}\phi_{zA}%
-\beta_{n}A_{bz_{n}}\right)  ~, \tag{A9}%
\end{equation}
and along $A_{ax}$,
\begin{equation}
u_{ax}[z,A]=c^{2}\delta_{ab}\frac{\delta\phi_{zA}}{\delta A_{bx}}~, \tag{A10}%
\end{equation}
where
\begin{equation}
\phi_{zA}=\varphi_{zA}-\eta\log\rho_{zA}^{1/2}~. \tag{A11}%
\end{equation}

\subsubsection*{Acknowledgements}

I would like to acknowledge Connor Dolan for insightful discussions at the
early stages of this project.

\end{document}